\documentclass[aps,prl,reprint,showpacs,superscriptaddress,longbibliography]{revtex4-1}
\usepackage[dvipdfmx]{graphicx}

\usepackage{amsmath,amssymb,amsfonts}
\usepackage{comment}
\usepackage{color}
\usepackage[T1]{fontenc}
\usepackage{textcomp}
\usepackage{hyperref}
\hypersetup{
setpagesize=false,
colorlinks=true,
linkcolor=blue,
citecolor=red,
}
\usepackage{bm}
\usepackage{physics}
\makeatletter
\let\MYcaption\@makecaption
\makeatother
\usepackage{subcaption}
\makeatletter
\let\@makecaption\MYcaption
\makeatother

\begin{document}

\title{Valley hydrodynamics in gapped graphene}

\author{Ryotaro Sano}
\email{sano.ryotaro.52v@st.kyoto-u.ac.jp}
\affiliation{%
Department of Physics, Kyoto University, Kyoto 606-8502, Japan
}%

\author{Daigo Oue}
\affiliation{%
Department of Physics, Kyoto University, Kyoto 606-8502, Japan
}%
\affiliation{%
The Blackett Laboratory, Department of Physics, Imperial College London, Prince Consort Road, Kensington, London SW7 2AZ, United Kingdom
}%
\affiliation{%
Kavli Institute for Theoretical Sciences, University of Chinese Academy of Sciences, Beijing, 100190, China.
}%

\author{Mamoru Matsuo}

\affiliation{%
Kavli Institute for Theoretical Sciences, University of Chinese Academy of Sciences, Beijing, 100190, China.
}%

\affiliation{%
CAS Center for Excellence in Topological Quantum Computation, University of Chinese Academy of Sciences, Beijing 100190, China
}%
\affiliation{%
RIKEN Center for Emergent Matter Science (CEMS), Wako, Saitama 351-0198, Japan
}%
\affiliation{%
Advanced Science Research Center, Japan Atomic Energy Agency, Tokai, 319-1195, Japan
}%

\date{\today}

\begin{abstract}
Recent experiments have revealed that novel nonequilibrium states consistent with the hydrodynamic description of electrons are realized in ultrapure graphene, which hosts the valley degrees of freedom. Here, we formulate a theory of electron hydrodynamics including dissipation processes of the valley angular momentum by employing the concept of micropolar fluids. As a result, our theory proposes a novel strategy to generate a valley polarization by the microrotation. We uncover that the rotational viscosity induces longitudinal valley currents which are second order in electric fields.
\end{abstract}

\maketitle
{\it Introduction.---}
The investigation of internal quantum degrees of freedom (DOF) of electrons in solids lies at the heart of condensed matter physics. The most-studied example is that of the electron spin, which gives rise to the vast field of spintronics with an eye on their potential for future electronics~\cite{spintronics_review,maekawa2017spin}. The advent of novel 2D materials which support massive Dirac fermions, exemplified by gapped graphene and transition metal dichalcogenides, has triggered research on alternative future electronics~\cite{Xiao2007-yy,Yao2008-we,DiXiao2012,Butler2013,Nguyen_2016,Liu2017,CHOI2017116}. In these systems, two inequivalent valleys $K$ and $-K$ reside at the corners of the hexagonal Brillouin zone. Similar to the spin, the valley labeling constitutes a discrete angular momentum for low energy carriers. From this point of view, the valley DOF has a potential use for information carriers, giving rise to an active research field called valleytronics as a promising concept for the next-generation electronics~\cite{Rycerz2007,Radisavljevic2011,Wang2012,Xu2014,Zhang2014,Schaibley2016,Vitale2018,Liu2019}.
Especially, the valley polarization, a nonequilibrium charge carrier imbalance between valleys, is the key to create valleytronic devices~\cite{Cao2012-sn,Zeng2012,Mak2012,Mak2016,Ye2017,Mak2018,Li2014-it,MacNeill2015-ca,Srivastava2015,Aivazian2015,Back2017PRL,Mak2014,doi:10.1126/science.1254966,Ye2016}. Therefore, a necessary requirement for valleytronic applications is the ability to generate and control the valley polarization.

Ultraclean 2D materials have also boosted the study of electron hydrodynamics, where electron-electron collisions are the dominant scattering processes. In fact, many pieces of clear evidence for hydrodynamic behaviors of charge carriers have been provided on graphene~\cite{Crossno2016,Bandurin2016,Bandurin2018,KrishnaKumar2017,Pellegrino2017,Berdyugin2019,Sulpizio2019,Gallagher2019,Polini2020,Ku2020,jenkins2020imaging,geurs2020rectification,Samaddar2021,kumar2021imaging}. These nonequilibrium behaviors of interacting systems close to equilibrium are well described by tracking the evolution of conserved quantities including internal microscopic DOF. However, the construction of hydrodynamics which deals with quantum DOF of electrons such as spin and valley is a nontrivial task. This is because dissipation processes of such an angular momentum cannot be captured by those of the vorticity, which follow directly from the momentum conservation law~\footnote{More detailed discussion is given in the supplemental materials}. Thus, electron hydrodynamics reaches the stage of going beyond the conventional Navier-Stokes equation.

\begin{figure}[t]
	\centering
	\includegraphics[width=0.8\columnwidth]{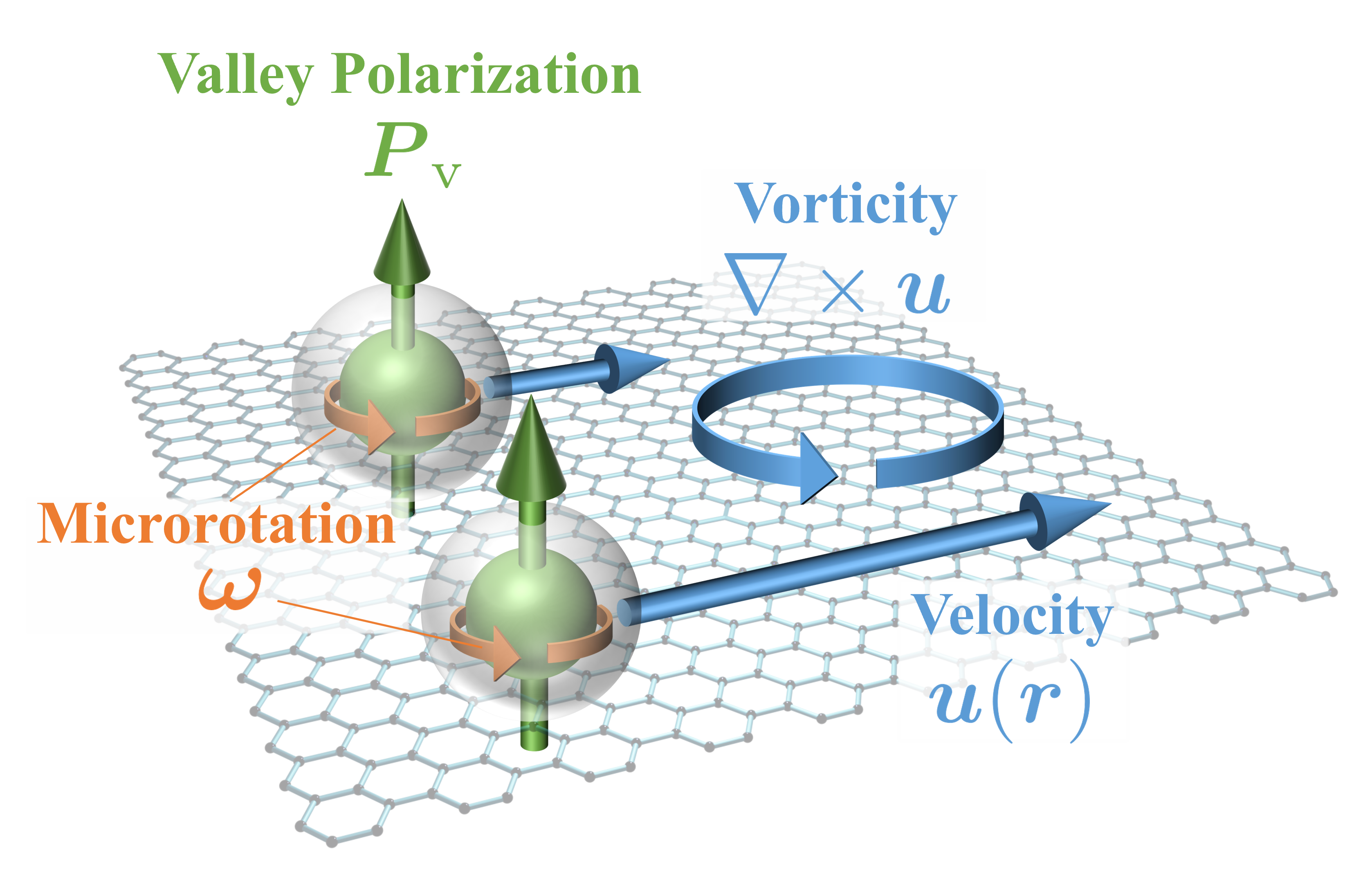}
	\caption{Schematics of the microrotation in valley hydrodynamics. The non-uniform velocity $\vb*{u}(\vb*{r})$ gives rise to the vorticity $\nabla\times\vb*{u}$. On the other hand, the microrotation $\vb*{\omega}$ is the internal angular momentum stems from the valley polarization $P_\mathrm{v}$ of fluid elements. The microrotation relaxes towards the vorticity due to the rotational viscosity.}
	\label{fig1}
\end{figure}

The concept of micropolar fluids gives a new twist to the previous studies of electron hydrodynamics. Micropolar fluid is an extended fluid with an internal rotation of fluid elements so-called the microrotation [see Fig.\ref{fig1}]. In micropolar fluids, the antisymmetric components of the stress tensor play a crucial role in dissipation processes of angular momentum, thereby switching on the rotational viscosity between the vorticity and the microrotation~\cite{Eringen1966,Eringen_1999,eringen2001microcontinuum,Lukaszewicz_1999}. These features have led to special attention in micropolar fluids due to their many applications in liquid crystals~\cite{chandrasekhar1992liquid,deGennes1993physics}, ferrofluids~\cite{Odenbach2002,Shliomis1972,rosensweig2013ferrohydrodynamics}, spintronics~\cite{Takahashi2016}, and active matter~\cite{Marchetti2013,Markovich_2019,Markovich2021,Shaik2021-th}. Furthermore, the phenomenon of angular momentum conversion between internal DOF of quantum particles and mechanical rotation have attracted great interest in various fields, ranging from nuclear physics~\cite{Adamczyk2017,Florkowski2018,HATTORI2019100} to condensed matter physics~\cite{Matsuo2013,Matsuo2017,Kobayashi2017,Hughes2011}. This naturally motivates us to study electron hydrodynamics including internal quantum DOF, especially the valley angular momentum, with employing the framework of micropolar fluids. Our study proposes a new strategy for controlling the valley polarization by the microrotation and may open up various avenues for research on valley hydrodynamics.

\begin{figure}[t]
		\centering
        \begin{subfigure}{0.9\columnwidth}
        \centering
        \includegraphics[width=0.9\columnwidth]{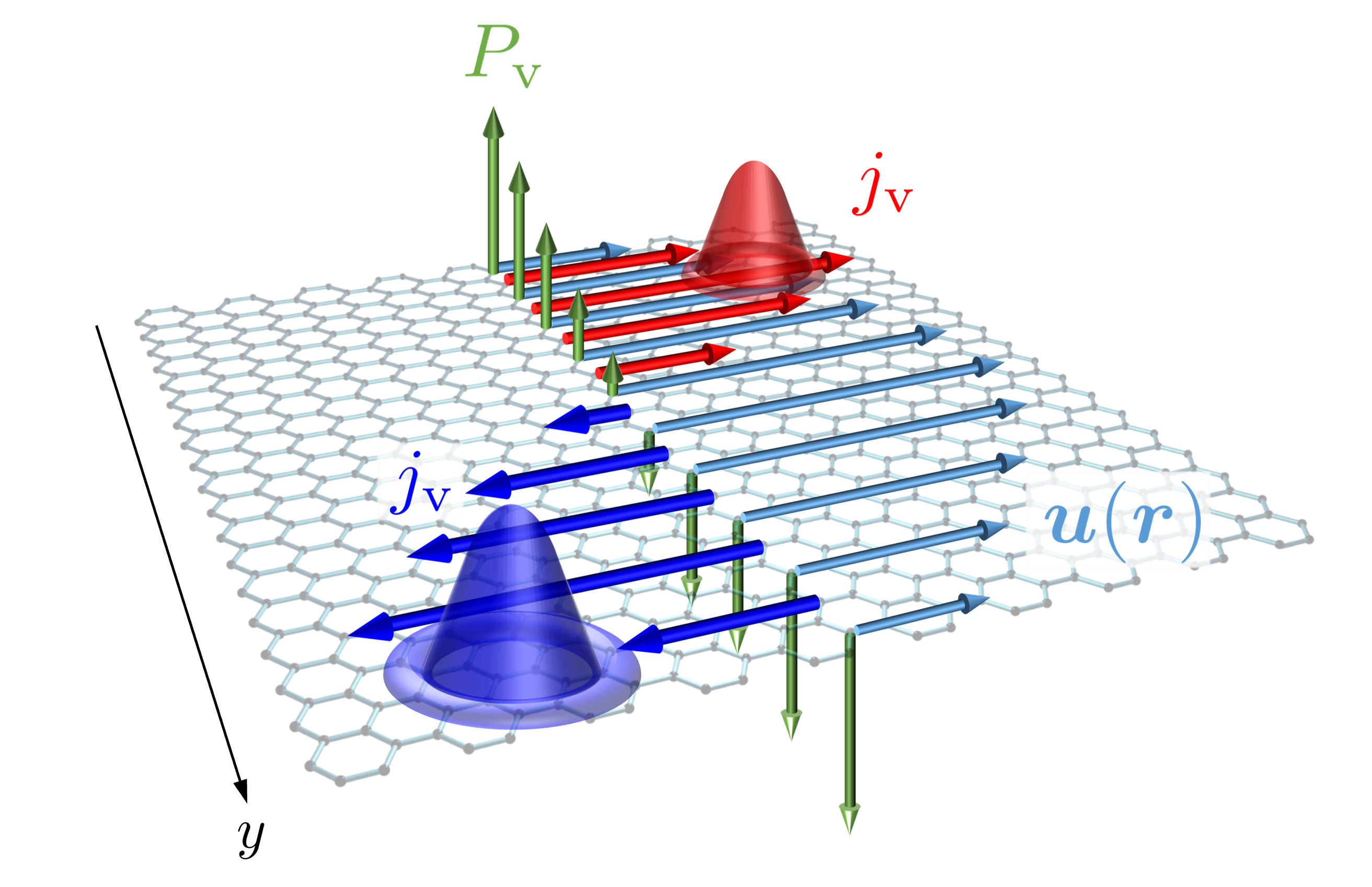}
        \subcaption{Valley hydrodynamic generation}
        \label{fig:valleypicture}
        \end{subfigure}\\
        \begin{subfigure}{0.45\columnwidth}
        \centering
        \includegraphics[width=\columnwidth]{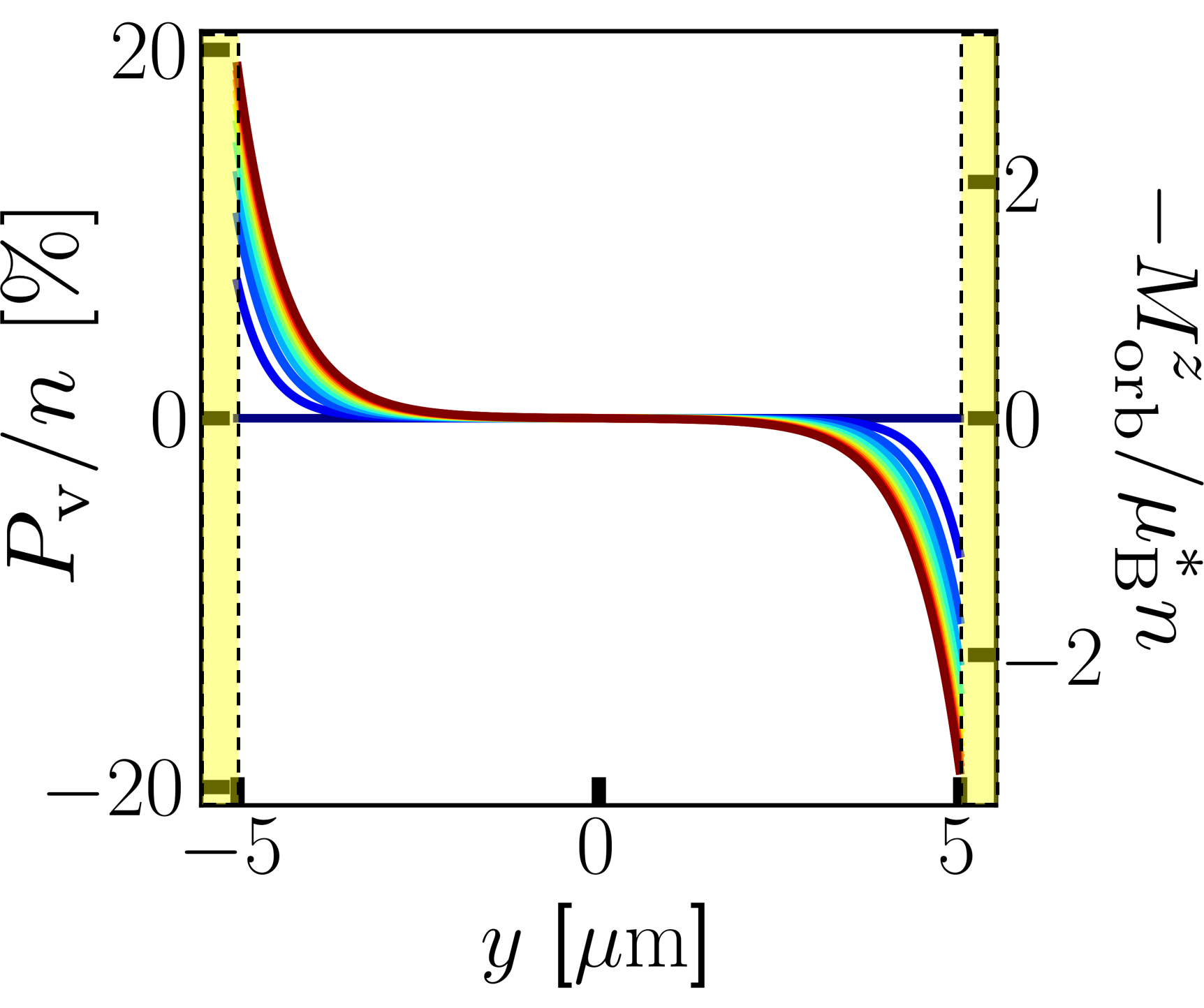}
        \subcaption{Valley polarization $P_\mathrm{v}$}
        \label{fig:valleypolarization}
        \end{subfigure}
        \hspace{3mm}
        \begin{subfigure}{0.48\columnwidth}
        \centering
        \includegraphics[width=\columnwidth]{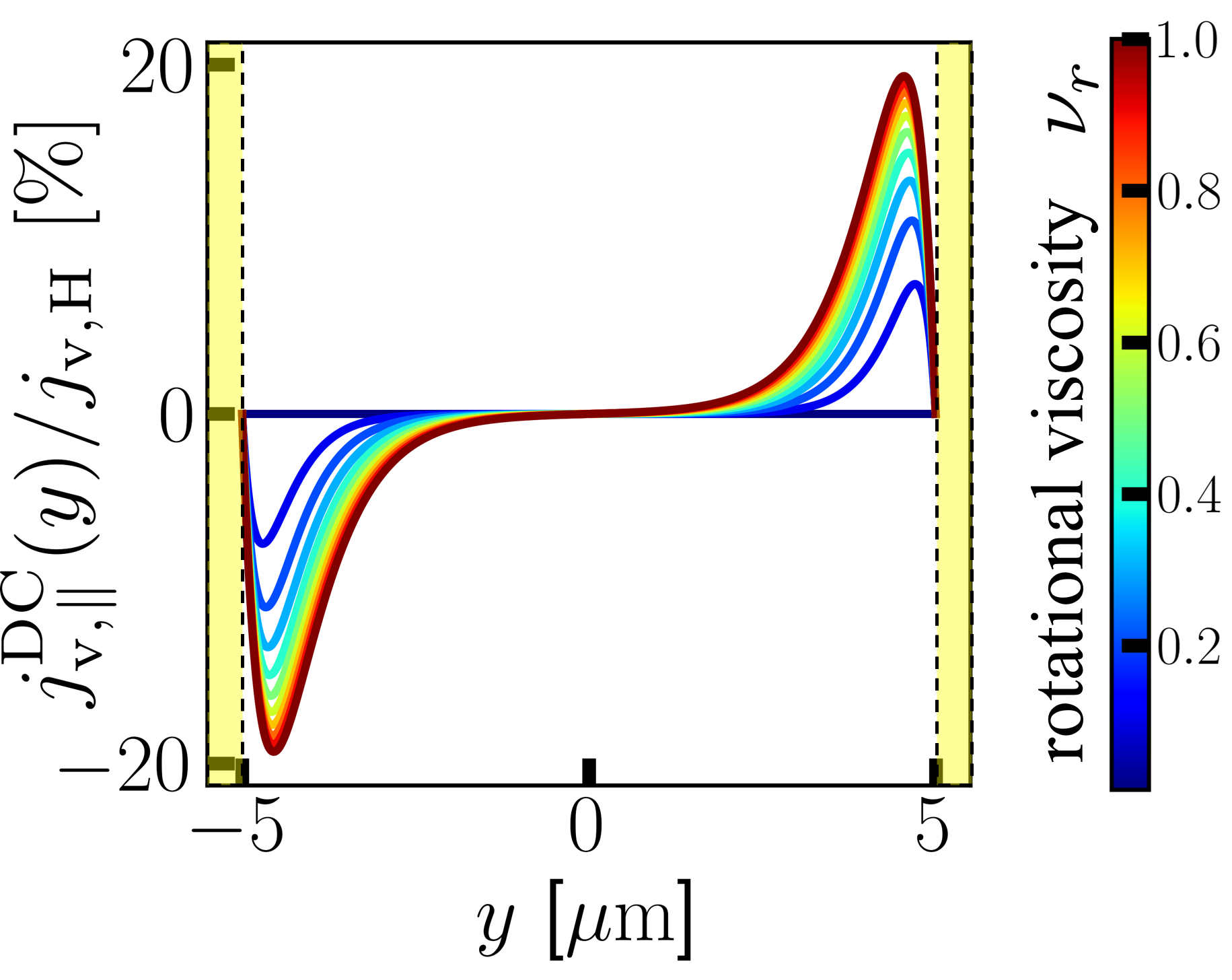}
        \subcaption{Valley current $j_\mathrm{v,\parallel}^\mathrm{DC}$}
        \label{fig:valleycurrent}
        \end{subfigure}
		\caption{(a) Schematics of the valley hydrodynamic generation. In hydrodynamic regime, the valley polarization $P_\mathrm{v}$ is induced by the microrotation via the valley-microrotation coupling with non-uniform electron velocity $\vb*{u}(\vb*{r})$, and the longitudinal valley current is generated as $\vb*{j}_\mathrm{v}=P_\mathrm{v}\vb*{u}$. Plot of (b) the valley polarization and (c) the valley current in the $y$-range [$-w/2,w/2$] under DC electric fields for several rotational viscosities. We use the parameters; $w=10\,\mu\mathrm{m}$, $E=1\times 10^6\,\mathrm{Vm^{-1}}$, $\nu=0.1\,\mathrm{m^2s^{-1}}$, $I\sim10^{-12}\,\mathrm{m^2}$, $\tau_\mathrm{R}=\tau_\mathrm{vf}=1\times 10^{-12}\,\mathrm{s}$, $a=2.46\,\mathrm{\mathring{A}}$, $t=2.82\,\mathrm{eV}$, $\Delta=0.28\,\mathrm{eV}$, $\mu=0.15\,\mathrm{eV}$, $n=1.4\times 10^{15}\,\mathrm{m^{-2}}$, $\rho=2.8\times10^{-17}\,\mathrm{kgm^{-2}}$, and the valley Hall current $j_{\mathrm{v},\mathrm{H}}=2.76\times10^{21}\,\mathrm{m^{-1}s^{-1}}$.}
		\label{fig2}
	\end{figure}

\begin{figure}[t]
    \centering
    \includegraphics[width=0.72\columnwidth]{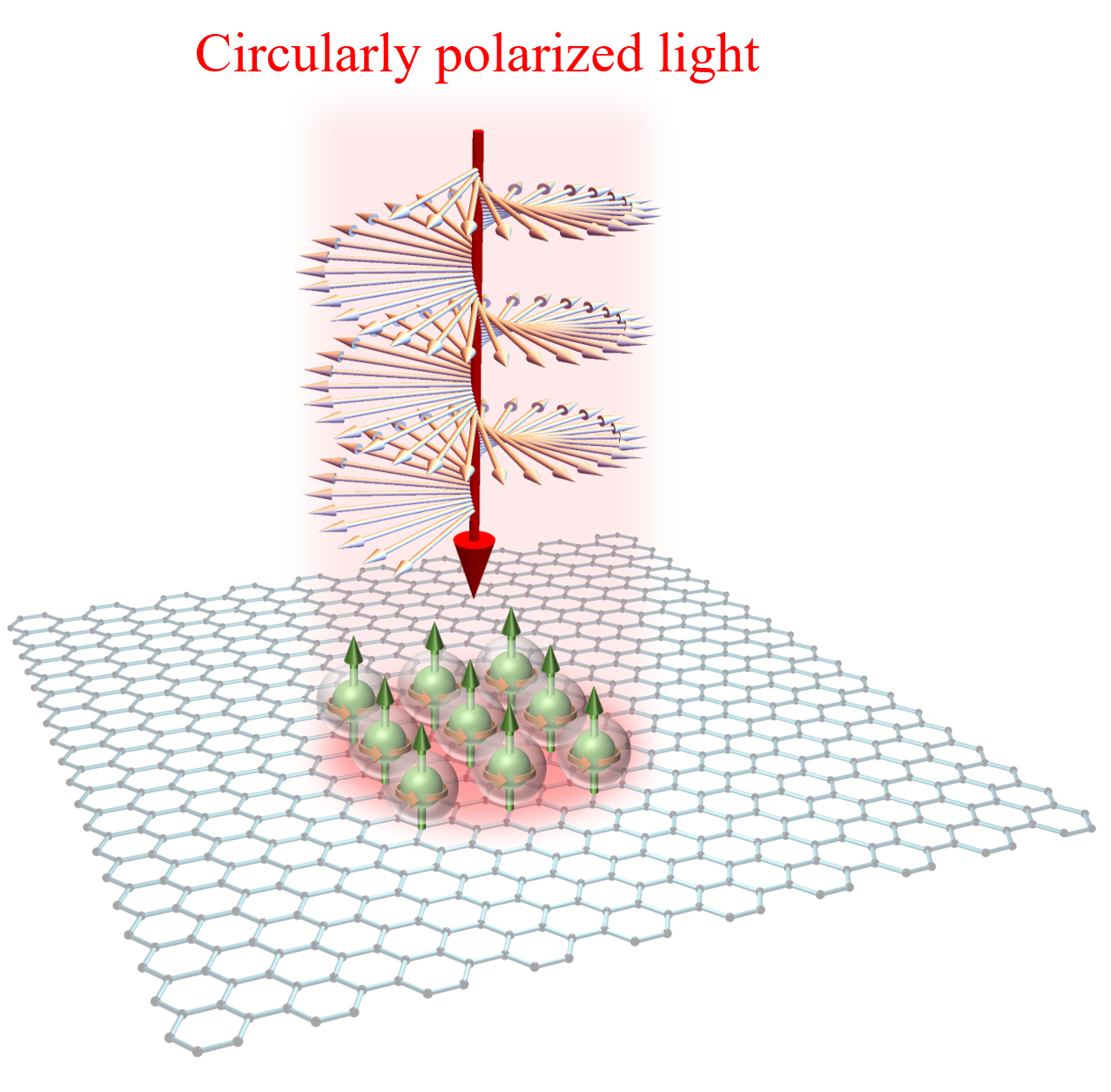}
    \caption{Schematics of circular photovalley generation. The microrotation is directly induced by irradiating CPL via the inverse Faraday effect. As a result, a DC valley polarization is generated.}
    \label{fig:CPL}
\end{figure}

In this Letter, we derive the hydrodynamic equations with the valley DOF for noncentrosymmetric 2D honeycomb lattice systems, which are correct up to the first order in the drift velocity and the microrotation. We identify the valley DOF as a microrotation and reveal that our theory acquires an emergent conservation law for the microrotation owing to the valley-microrotation coupling. From a symmetry viewpoint, we find that this interaction can appear only in the systems without inversion symmetry.
A key ingredient for valleytronics is a controllable way of population imbalance between the two valleys, thereby producing a valley polarization. While previous works showed that a valley polarization can be controlled by optical~\cite{Cao2012-sn,Zeng2012,Mak2012,Mak2016,Ye2017,Mak2018}, magnetic~\cite{Li2014-it,MacNeill2015-ca,Srivastava2015,Aivazian2015,Back2017PRL} and electrical~\cite{Mak2014,doi:10.1126/science.1254966,Ye2016} means. Our study shows a microfluidic approach to generate such a valley polarization by the microrotation. We also predict nonlinear valley dynamics including longitudinal valley currents [Fig.\ref{fig2}] and a circular dichroic valley polarization induced by off-resonant light [Fig.\ref{fig:CPL}].

{\it Formulation.---}
We outline how to derive the hydrodynamic equations for noncentrosymmetric graphene with a staggered sublattice potential. We start from the Boltzmann equation which governs the evolution of the electron distribution function $f_{\alpha\tau}$ for band $\alpha$ and valley $\tau$,
\begin{align}
    &\frac{\partial f_{\alpha\tau}}{\partial t}+\dot{\vb*{r}}_{\alpha\tau}\cdot\frac{\partial f_{\alpha\tau}}{\partial\vb*{r}}+\dot{\vb*{k}}_{\alpha\tau}\cdot\frac{\partial f_{\alpha\tau}}{\partial\vb*{k}}\nonumber\\
    &=-\frac{f_{\alpha\tau}-f_{\alpha\tau}^{\mathrm{N}}}{\tau_{\mathrm{N}}}-\frac{f_{\alpha\tau}-f_0}{\tau_{\mathrm{R}}}-\frac{f_{\alpha\tau}-f_{\alpha-\tau}}{\tau_{\mathrm{vf}}},
\end{align}
where $f_0$ is the Fermi-Dirac (global equilibrium) distribution function. $\tau_\mathrm{N}$, $\tau_\mathrm{R}$ and $\tau_{\mathrm{vf}}$ are the relaxation times for normal\,(N), resistive\,(R) and valley flipping processes. Here, N process conserves the linear momentum, while R process does not. If we construct an electron wave packet near the valley center, the semiclassical equations of motion read~\cite{DiXiao2010review}
\begin{align}
    \dot{\vb*{r}}_{\alpha\tau}=\frac{1}{\hbar}\frac{\partial\epsilon_{\alpha\tau}}{\partial\vb*{k}}-\dot{\vb*{k}}_{\alpha\tau}\times\vb*{\Omega}_{\alpha\tau},\quad\dot{\vb*{k}}_{\alpha\tau}=-\frac{e}{\hbar}\vb*{E},
\end{align}
where electric fields $\vb*{E}$ can be time-dependent. $\epsilon_{\alpha\tau}(\vb*{k})$ and $\vb*{\Omega}_{\alpha\tau}(\vb*{k})$ are the band energy and the Berry curvature of the Bloch electrons respectively. Due to the lack of inversion symmetry, $\vb*{\Omega}_{\alpha\tau}$ is allowed to have nonzero values for any $\vb*{k}$.

Following the standard approach~\cite{Landau_Kinetics,Lucas_2018,NAROZHNY2019167979,Narozhny2019,Narozhny2021,Kiselev2020}, the continuity equations for the carrier density and the linear momentum are obtained as follows:
\begin{gather}
    \frac{\partial n}{\partial t}+\nabla\cdot\vb*{j}=0,\label{particle0}\\
    \frac{\partial P_i}{\partial t}+\frac{\partial\Pi_{ij}}{\partial x_j}=-enE_i-\frac{P_i}{\tau_{\mathrm{R}}},\label{momentum0}
\end{gather}
where $n$ and $\vb*{P}$ are the carrier and the linear momentum densities respectively, $-en\vb*{E}$ is the driving force due to external electric fields. In Eqs.\eqref{particle0} and \eqref{momentum0}, $\vb*{j}$ and $\Pi_{ij}$ are the corresponding fluxes of each density.

A monolayer graphene with a staggered sublattice potential breaking the inversion symmetry is a concrete example for considering valley hydrodynamics. Staggered sublattice potential is generally expected in epitaxial graphene on SiC substrates~\cite{Berger2004,Zhou2007,Tromp2009,Hupalo2009,Jariwala2011,Jeon2013,Nevius2015}. The effective Hamiltonian describing electron states in the vicinity of the $K$ and $-K$ points is given by~\cite{Xiao2007-yy}
\begin{equation}
    H_\tau=at(\tau k_x\sigma_x+k_y\sigma_y)+\frac{\Delta}{2}\sigma_z,\label{hamiltonian}
\end{equation}
where $a$ and $t$ are the lattice constant and the nearest-neighbor hopping parameter, $\vb*{k}=(k_x,k_y)$ are the two components of the wave vector measured from the valley center, $\vb*{\sigma}$'s are the Pauli matrices representing a pseudospin from the sublattice DOF, and $\tau=\pm1$ is the valley index labeling the two inequivalent valleys. Note that its band structure $\epsilon_{\alpha\tau}(\vb*{k})$ has no dependence on the valley, while the Berry curvature $\Omega_{\alpha\tau}^z(\vb*{k})$ has a valley-contrasting property. Because of large separation of two valleys in the momentum space, intervalley scatterings are strongly suppressed~\cite{intervalley_scattering1,intervalley_scattering2,intervalley_scattering3}, implying the potential for regarding the valley polarization as a conserved quantity. Therefore, the effective hydrodynamic theory acquires an emergent continuity equation for the valley polarization:
\begin{equation}
    \frac{\partial P_{\mathrm{v}}}{\partial t}+\nabla\cdot\vb*{j}_{\mathrm{v}}=-\frac{P_{\mathrm{v}}}{\tau_{\mathrm{R}}}-\frac{2P_{\mathrm{v}}}{\tau_{\mathrm{vf}}},\label{valley}
\end{equation}
where $P_{\mathrm{v}}$ and $\vb*{j}_{\mathrm{v}}$ are the valley polarization and the valley current. Here, we have defined the valley polarization $P_{\mathrm{v}}\equiv n_{K}-n_{-K}$ as a population imbalance between the two valleys in analogy to the spin polarization. We should note that not only valley flipping processes but also R process contribute to the relaxation of the valley polarization. This indicates that the valley DOF combines a linear momentum and an angular momentum.

In hydrodynamic regime, $\tau_\mathrm{N}\ll\tau_\mathrm{R},\tau_\mathrm{vf}$, the system reaches local equilibrium via N electron-electron scatterings, which conserve both the linear momentum and the valley polarization of the electron system. For this reason, we assume that the distribution functions are described as
\begin{equation}
    f^{\mathrm{N}}_{\alpha\tau}=\left[\exp\left(\frac{\epsilon_{\alpha\tau}-\hbar\vb*{k}\cdot\vb*{u}-\tau\hbar\omega_z-\mu}{k_\mathrm{B}T}\right)+1\right]^{-1},
\end{equation}
which is referred to as the local equilibrium distribution function. Here, the drift velocity $\vb*{u}$ and the microrotation $\omega_z$ are corresponding parameters for conserved quantities $\vb*{P}$ and $P_\mathrm{v}$. From a symmetry viewpoint, the absence of inversion symmetry allows for the interplay between the valley DOF and an angular momentum; examples include the spin-valley coupling~\cite{DiXiao2012,Sanchez2016} and the valley-vorticity coupling~\cite{ominato2021valley}. Here, the microrotation is the internal angular momentum of the fluid elements and we referred to $\tau\omega_z$ as the valley-microrotation coupling.

Since the relevant conduction and valence bands are well described by Eq.\eqref{hamiltonian} for low doping level, we use a quadratic dispersion $\epsilon_{\alpha\tau}=\alpha[\Delta/2+\hbar^2\vb*{k}^2/2m^\ast]$ with an effective mass $m^\ast\equiv\hbar^2\Delta/2a^2t^2$ in the vicinity of the $K$ and $-K$ points in the following analysis. Under this assumption, we obtain the valley polarization and the valley current in terms of hydrodynamic variables:
\begin{align}
    P_{\mathrm{v}}&=\hbar\omega_z\sum_{\alpha,\tau}\int[\mathrm{d}\vb*{k}]\left(-\frac{\partial f_0}{\partial\epsilon}\right)\simeq\hbar\omega_zD(\mu),\label{pv}\\
    \vb*{j}_{\mathrm{v}}&=P_{\mathrm{v}}\vb*{u}+\frac{e}{\hbar}\vb*{E}\times\sum_{\alpha,\tau}\tau\int[\mathrm{d}\vb*{k}]\vb*{\Omega}_{\alpha\tau}f_0,\label{jv}
\end{align}
with the density of states $D(\epsilon)$ and $\int[\mathrm{d}\vb*{k}]\equiv\int\mathrm{d}\vb*{k}/(2\pi)^2$. These results indicate that the microrotation leads to a valley polarization. The second term in Eq.~\eqref{jv} is the well-known valley Hall effect~\cite{Xiao2007-yy}, on the other hand, the first term is the longitudinal valley current which is one of our main results.

Combining the valley polarization conservation law Eq.~\eqref{valley} with Eqs.~\eqref{pv} and~\eqref{jv}, we end up with the hydrodynamic equations:
\begin{gather}
    \frac{\partial\vb*{u}}{\partial t}+(\vb*{u}\cdot\nabla)\vb*{u}+\frac{1}{\rho}\nabla p=-\frac{en\vb*{E}}{\rho}-\frac{\vb*{u}}{\tau_\mathrm{R}},\label{Euler}\\
    \frac{\partial\omega_z}{\partial t}+(\vb*{u}\cdot\nabla)\omega_z+\omega_z(\nabla\cdot\vb*{u})=-\frac{\omega_z}{\tau_{\mathrm{inter}}},\label{angular}
\end{gather}
where $\rho$ is the mass density and $\tau_{\mathrm{inter}}=(1/\tau_{\mathrm{R}}+2/\tau_\mathrm{vf})^{-1}$ is the relaxation time for intervalley scatterings. Notably, our theory acquires an emergent conservation law for the microrotation and therefore it has a close analogy to micropolar fluids.

In the following analysis, we assume that the fluid is incompressible. We also phenomelogically introduce the rotational viscous torques proportional to the deviation of the microrotation $\vb*{\omega}$ from the vorticity $\nabla\times\vb*{u}/2$. According to Refs.~\cite{Eringen1966,Eringen_1999,eringen2001microcontinuum,Lukaszewicz_1999}, the equations governing the flow of incompressible micropolar fluids are given by
\begin{gather}
    \frac{D\vb*{u}}{Dt}=(\nu+\nu_r)\Delta\vb*{u}+2\nu_r\nabla\times\vb*{\omega}-\frac{en\vb*{E}}{\rho}-\frac{\vb*{u}}{\tau_{\mathrm{R}}},\label{momentum}\\
    I\frac{D\omega_z}{Dt}=2\nu_r[(\nabla\times\vb*{u})_z-2\omega_z]-I\frac{\omega_z}{\tau_\mathrm{inter}},\label{microrot}
\end{gather}
where $D/Dt=\partial/\partial t+\vb*{u}\cdot\nabla$ is the convective derivative. $\nu$ and $\nu_r$ are the kinematic and the rotational viscosities respectively. Here we have introduced the microinertia $I$, which is moment of inertia per unit fluid mass. In valley hydrodynamics, the motion of electron fluids is described by the drift velocity $\vb*{u}$ and the microrotation $\vb*{\omega}$.

{\it Orbital magnetization.---}
The orbital magnetic moment is one of the valley-contrasting parameters in 2D materials~\cite{Xiao2007-yy,Xu2014}. In gapped graphene, the orbital magnetic moment is given by
\begin{equation}
    \mathfrak{m}_{\alpha\tau}^z(\vb*{k})=-\tau\frac{e}{2\hbar}\frac{2a^2t^2\Delta}{\Delta^2+4a^2t^2k^2},
\end{equation}
which can be intuitively interpreted as the self-rotation of the Bloch wave packet.
Its value at the valley points has a suggestive form: $\mathfrak{m}_{\alpha\tau}^z(0)=-\tau\mu_\mathrm{B}^\ast$, where $\mu^\ast_\mathrm{B}\equiv e\hbar/2m^\ast$ resembles the Bohr magneton for the electron spin. Because $\mu^\ast_\mathrm{B}$ is about 40 times larger than the Bohr magneton, the response to a perpendicular magnetic field is dominated by the orbital magnetic moment. The orbital magnetization consists of the orbital moment of carriers plus a correction from the Berry curvature~\cite{Thonhauser2011}:
\begin{align}
    &M^z_\mathrm{orb}=\sum_{\alpha,\tau}\int[\mathrm{d}\vb*{k}]\mathfrak{m}^z_{\alpha\tau}f_{\alpha\tau}+\frac{1}{\beta}\frac{e}{\hbar}\sum_{\alpha,\tau}\int[\mathrm{d}\vb*{k}]\nonumber\\
    &\times\Omega^z_{\alpha\tau}\log[1+e^{-\beta(\epsilon_{\alpha\tau}-\hbar\vb*{k}\cdot\vb*{u}-\hbar\omega_z\tau-\mu)}].
\end{align}
After straightforward calculation, we obtain the orbital magnetization,
\begin{align}
    M^z_\mathrm{orb}&=\hbar\omega_z\sum_{\alpha,\tau}\tau\int[\mathrm{d}\vb*{k}]\left[\mathfrak{m}_{\alpha\tau}^z\left(-\frac{\partial f_0}{\partial\epsilon}\right)+\frac{e}{\hbar}\Omega_{\alpha\tau}^zf_0\right].\label{OM}
\end{align}
This result also supports that the microrotation has a meaning of an angular momentum. The spatial profile of the orbital magnetization can be detected with magneto-optical Kerr rotation microscopy~\cite{Mak2014,Kato2004,Lee2016,Lee2017}. By using this experimental setup, a population difference in the two valleys can be also detected as a signal of the orbital magnetization [see Fig.~\ref{fig:valleypolarization}, as a demonstration].

{\it Valley hydrodynamic generation.---}
The most significant consequences of our theory is that the interplay between the valley-microrotation coupling and the viscous effects gives rise to an unprecedented longitudinal nonlinear valley current in finite size systems, which has not been addressed so far. We consider the Poiseuille flow in gapped graphene with finite width $w$ in the $y$-direction, which most clearly characterizes the hydrodynamic transport. When we apply DC electric fields in the $x$-direction and take no-slip boundary conditions $u_x(\pm w/2)=0$, the electron fluids form the Poiseuille flow with the velocity profile given by
\begin{equation}
    u_x(y)=-\frac{en\tau_{\mathrm{R}}}{\rho}\left[1-\frac{\cosh(y/\ell)}{\cosh(w/2\ell)}\right]E.
\end{equation}
The microrotation is also calculated as
\begin{equation}
    \omega_z(y)=-\frac{\tau_\mathrm{eff}}{\tau_r}\frac{1}{2}\frac{\partial u_x}{\partial y}=-\frac{en\tau_{\mathrm{R}}}{2\rho\ell}\frac{\tau_\mathrm{eff}}{\tau_r}\frac{\sinh(y/\ell)}{\cosh(w/2\ell)}E,
\end{equation}
where
\begin{equation}
\ell\equiv\sqrt{\left(\nu+\nu_r\frac{\tau_r}{\tau_r+\tau_\mathrm{inter}}\right)\tau_{\mathrm{R}}},
\end{equation}
is a characteristic length that determines the scale of viscous effects. Here, $\tau_r^{-1}\equiv4\nu_r/I$ and $\tau_\mathrm{eff}\equiv(1/\tau_\mathrm{inter}+1/\tau_r)^{-1}$ are the rotational and effective relaxation times. From Eq.\eqref{jv}, we obtain the longitudinal valley current profile as [Fig.\ref{fig:valleycurrent}]
\begin{align}
    j^{\mathrm{DC}}_{\mathrm{v},\parallel}(y)=&\left(\frac{en\tau_\mathrm{R}}{\rho}E\right)^2\frac{\hbar D(\mu)}{2\ell}\frac{\tau_\mathrm{eff}}{\tau_r}\nonumber\\
    &\times\frac{\sinh(y/\ell)}{\cosh(w/2\ell)}\left[1-\frac{\cosh(y/\ell)}{\cosh(w/2\ell)}\right].
\end{align}
We should note that the rotational viscosity $\nu_r$ is necessary for realizing a longitudinal nonlinear valley current under DC electric fields. Similar to DC electric fields, an AC electric field along the $x$-direction $E_x(t)=\Re[\tilde{E}e^{-i\Omega t}]$ also induces the Poiseuille flow and leads to the same solutions:
\begin{equation}
    \tilde{u}_x(y,\Omega)=\frac{u_x(y)}{1-i\Omega\tau_\mathrm{R}},\quad \tilde{\omega}_z(y,\Omega)=\frac{\omega_z(y)}{(1-i\Omega\tau_\mathrm{eff})(1-i\Omega\tau_\mathrm{R})},
\end{equation}
except for the replacement of $\ell$ by
\begin{equation}
    \tilde{\ell}(\Omega)=\sqrt{\left(\nu+\nu_r\frac{\tau_r}{\tau_r+\tau_\mathrm{inter}}\frac{1-i\Omega\tau_\mathrm{inter}}{1-i\Omega\tau_\mathrm{eff}}\right)\frac{\tau_\mathrm{R}}{1-i\Omega\tau_{\mathrm{R}}}}.
\end{equation}
Because of the intrinsic nonlinearity of the longitudinal valley current, $\vb*{j}_\mathrm{v,\parallel}$ is composed of the valley counterparts of the rectification and the second harmonic generation:
\begin{equation}
    j_{\mathrm{v},\parallel}^\mathrm{AC}(y,t)=j_{\mathrm{v},\parallel}^0(y)+j_{\mathrm{v},\parallel}^{2\Omega}(y,t).
\end{equation}

{\it Circular photovalley generation.---}
The rotational viscosity-induced valley transport discussed above can be interpreted as a phenomenon of angular momentum conversion between the fluid vorticity and the valley DOF. From this viewpoint, we consider a different scenario for generating a valley polarization by circularly polarized light (CPL). CPL with the electric component $\vb*{E}(t)=E_0(\cos\Omega t,\xi\sin\Omega t)$ induces a circular motion of electrons, which in turn generates a DC orbital magnetization:
\begin{equation}
{M}_{\mathrm{orb}}^z=-\frac{ne^3}{4{m^\ast}^2\Omega^3}\xi E_0^2.\label{IFE}
\end{equation}
This phenomenon is known as the inverse Faraday effect~\cite{HERTEL2006L1,ZHANG2009L73,Battiato2014,Potashin2020} owing to the fact that CPL has a spin angular momentum proportional to $\xi E_0^2$~\cite{Barnett2010,Cameron2012,Bliokh2013,Bliokh2014,BLIOKH20151}. Here, the different chirality indices $\xi=\pm1$ correspond to the clockwise/counterclockwise circular polarizations. In our hydrodynamic formulation, the orbital magnetization is described as Eq.\eqref{OM}, therefore, the inverse Faraday effect can be regarded as a direct transfer mechanism of angular momentum from CPL to the microrotation.

We are now ready to discuss the generation of a valley polarization by CPL dubbed \textit{circular photovalley generation} [Fig.\ref{fig:CPL}]. In contrast to the above discussion, we consider bulk systems with normally-incident CPL.
We start from the hydrodynamic equations:
\begin{gather}
\frac{D\vb*{u}}{Dt}=(\nu+\nu_r)\Delta\vb*{u}+2\nu_r\nabla\times\vb*{\omega}-\frac{en\vb*{E}}{\rho}-\frac{\vb*{u}}{\tau_\mathrm{R}},\label{Eq:velocity}\\
\frac{D\omega_z}{Dt}=\frac{1}{\tau_r}\left[\frac{(\nabla\times\vb*{u})_z}{2}-\omega_z\right]-\frac{\omega_z}{\tau_\mathrm{inter}}+g\xi E_0^2,\label{Eq:micro}
\end{gather}
where an external angular momentum pumping term stems from the inverse Faraday effect is introduced. By solving Eqs.\eqref{Eq:velocity} and \eqref{Eq:micro}, the DC component of the microrotation in the second order in electric fields is obtained as
\begin{gather}
    \omega_z^0=\tau_\mathrm{eff}g\xi E_0^2,\label{Eq:micro2}
\end{gather}
giving rise to a nonlinear DC valley polarization:
\begin{gather}
    P_\mathrm{v}^0=\hbar D(\mu)\tau_\mathrm{eff}g\xi E_0^2.
\end{gather}
Here, the coefficient of $g$ is estimated as
\begin{equation}
    \tau_\mathrm{eff}g=\mathrm{sgn}(\mu)\frac{e^2}{4\hbar}\frac{a^2t^2}{(\hbar\Omega)^3}\frac{\mu^2-\Delta^2/4}{\Delta^2/4},
\end{equation}
in consistent with Eqs.\eqref{OM} and \eqref{IFE}.
In stark contrast to the previous works~\cite{Cao2012-sn,Zeng2012,Mak2012,Mak2016,Ye2017,Mak2018}, we do not rely on inter-band transition processes, therefore, on-resonant light is not required. This suggests that our hydrodynamic approach broadens the frequency range of CPL and also allows ultrafast manipulation of the valley polarization. This can be achieved by combining the valley DOF with the concept of micropolar fluids. Furthermore, the sign of the valley polarization can be tuned by the chirality $\xi$ of CPL, and hence circular dichroism appears in the valley polarization.

{\it Discussion.---}%
We propose an experimental setup how to determine $\nu_r$ below. In order to estimate the rotational viscosity experimentally, valley injection provides a reasonable measure. When we inject valley current from the proximity valley Hall materials into valley hydrodynamic materials, the induced valley polarization leads to the non-uniform microrotation profile $\vb*{\omega}(\vb*{r})$ due to the valley-microrotation coupling. Then, the fluid velocity $\vb*{u}$ is generated by the term $\nu_r\nabla\times\vb*{\omega}$ in Eq.\eqref{momentum}. Therefore, we can estimate the rotational viscosity from the observed velocity profile.

\begin{table}[t]
\caption{Comparison of the valley hydrodynamic generation (VHG) and circular photovalley generation (CPVG) as phenomena of angular momentum conversion via valley-microrotation coupling. System, source field, and mechanism are summarized.}
    \centering
    \begin{tabular}{ccc}
    \hline
    {} & VHG & CPVG \\
    \hline \hline
    System & Finite size & Bulk \\
    Source & Electric field & Circularly polarized light\\
    Mechanism & Rotational viscosity & Inverse Faraday effect \\
    \hline
    \end{tabular}
    \label{tab:valleytransport}
\end{table}

{\it Conclusion.---}%
In summary, we have developed a basic framework of valley hydrodynamics in noncentrosymmetric graphene with a staggered sublattice potential, which is composed of the Euler equation Eq.\eqref{Euler} and the balance equation for the microrotation Eq.\eqref{angular}. In addition, we have investigated the interplay between the valley DOF and the microrotation, and elucidated that the valley polarization can be controlled by the microrotation. Our hydrodynamic theory also reveals nonlinear valley dynamics. For example, the rotational viscosity $\nu_r$ provides a longitudinal nonlinear valley current, which gives rise to the valley counterparts of the rectification and the second harmonic generation. As discussed, $\nu_r$ can be measured by valley-induced hydrodynamic flow generation. Furthermore, the concept of micropolar fluids sheds light on a rich physics of angular momentum conversion, exemplified by a circular dichroic valley polarization induced by off-resonant light. These results are summarized in Table \ref{tab:valleytransport}.

The conventional strategy for designing electronic devices in spintronics or valleytronics has been creating confined nanostructure in order to achieve functionality. On the other hand, in hydrodynamic regime, the flow of electrons can become spatially non-uniform due to the viscosities even when the material structure is homogeneous. This suggests a new design guideline for innovative device functionality without nanostructure. Therefore, we believe that the present results provide a building block for future electronics and will pave the way to valleytronic applications of electron hydrodynamics.

\begin{acknowledgements}
The authors are grateful to Yuya Ominato, Shin Kaneshiro, Riki Toshio, Koki Shinada, Hideaki Nishikawa and Hisao Hayakawa for valuable discussions. R.S. thanks Satoshi Kusaba, Kohei Nagai and Kento Uchida for providing helpful comments from an experimental point of view. D.O. is funded by the President’s PhD Scholarships at Imperial College London.
This work was supported by the Priority Program of Chinese Academy of Sciences under Grant No. XDB28000000, and by JSPS KAKENHI for Grants (Nos. 20H01863 and 21H04565) from MEXT, Japan.
\end{acknowledgements}


\bibliography{ref}

\clearpage

\onecolumngrid
\begin{center}
{\large {\bfseries Supplemental Materials for \\ ``Valley hydrodynamics in gapped graphene''}}
\end{center}
\vspace{10pt}


\renewcommand{\theequation}{S.\arabic{equation}}
\setcounter{equation}{0}

\section{Beyond the Navier-Stokes equation}
In this section, we emphasize the need for introducing the concept of micropolar fluids. First, we see the failure of capturing the angular momentum dissipation from the momentum conservation law. In the conventional hydrodynamics, the momentum conservation law leads the well-known Navier-Stokes equation~\cite{Landau_Kinetics}:
\begin{equation}
   \frac{D\vb*{u}}{Dt}=\nu\laplacian\vb*{u}-\frac{\vb*{u}}{\tau_\mathrm{R}},\label{NSE}
\end{equation}
where $\vb*{u}$ is the fluid velocity and $D/Dt\equiv\partial/\partial t+\vb*{u}\vdot\grad$. The dissipation processes of linear momentum are represented by the kinematic viscosity term $\nu\laplacian\vb*{u}$ and the relaxation term $-\vb*{u}/\tau_\mathrm{R}$ in the right-hand side of Eq.\eqref{NSE}. Here, $\tau_\mathrm{R}$ is the momentum relaxation time. In order to consider angular momentum flow in the fluid, we take the curl of Eq.\eqref{NSE} and obtain an equation for the vorticity $\vb*{\Omega}\equiv\frac{1}{2}\curl\vb*{u}$:
\begin{equation}
    \frac{D\vb*{\Omega}}{Dt}=(\vb*{\Omega}\vdot\grad)\vb*{u}-\nu\laplacian\vb*{\Omega}-\frac{\vb*{\Omega}}{\tau_\mathrm{R}}.\label{Vorticity}
\end{equation}
Here, we have used the following identity:
\begin{equation}
    (\vb*{u}\vdot\grad)\vb*{u}=\grad\frac{1}{2}\vb*{u}^2-\vb*{u}\times(\curl\vb*{u})=\grad\frac{1}{2}\vb*{u}^2-\vb*{u}\times2\vb*{\Omega}.\nonumber
\end{equation}
The vorticity equation Eq.\eqref{Vorticity} is merely a mathematical transformation of Eq.\eqref{NSE}; therefore, Eq.\eqref{Vorticity} does not contain the information of dissipation processes of angular momentum.

In order to take the angular momentum dissipation into account, we go back to the fundamental level of the conservation laws. Let us consider angular momentum conservation by introducing the normal stress $\vb*{t}_n$ as well as the body torque $\vb*{l}$ and the couple stress $\vb*{\lambda}_n$:
\begin{equation}
    \dv{t}\int_{V(t)}\rho[\vb*{r}\times\vb*{u}+\vb*{s}]=\int_{\partial V(t)}(\vb*{r}\times\vb*{t}_n+\vb*{\lambda}_n)+\int_{V(t)}(\vb*{r}\times\rho\vb*{b}+\rho\vb*{l}),\label{AMC}
\end{equation}
where $\rho$ is the mass density, $\vb*{s}$ is the internal angular momentum per unit fluid mass, $\vb*{b}$ is the body force, $V(t)$ is the element of volume, and $\partial V(t)$ is the surface of the volume. By using the relations $\vb*{t}_n=\vb*{n}\vdot\vb{T}$ and $\vb*{\lambda}_n=\vb*{n}\vdot\vb{\Lambda}$, the right-hand side of Eq.\eqref{AMC} reads
\begin{equation}
    \int_{V(t)}(\vb*{r}\times\rho\vb*{b}+\rho\vb*{l}+\div\vb{\Lambda}+\vb*{r}\times(\div\vb{T})+\vb{T}_x),\label{AMC2}
\end{equation}
where $\vb{\Lambda}$ is the couple stress tensor and $(\vb{T}_x)_i=\epsilon_{ijk}T_{jk}$ is the axial vector component of the stress tensor $\vb{T}$. From Eqs.\eqref{AMC} and \eqref{AMC2}, we obtain
\begin{equation}
    \rho\vb*{r}\times\left(\frac{D\vb*{u}}{Dt}-\div\vb{T}-\vb*{b}\right)+\frac{D\vb*{s}}{Dt}=\rho\vb*{l}+\div\vb{\Lambda}+\vb{T}_x,\label{AMC3}
\end{equation}
where the bracket in the left-hand side is the momentum conservation law~\cite{Landau_Kinetics}:
\begin{equation}
    \dv{t}\int_{V(t)}\rho\vb*{u}=\int_{\partial V(t)}\vb*{t}_n+\int_{V(t)}\rho\vb*{b},\label{LMC}
\end{equation}
and hence vanishes. In the following, we consider isotropic micropolar fluids, whose internal angular momentum density can be described as
\begin{equation}
    \rho\vb*{s}=\rho I\vb*{\omega},\label{microinertia}
\end{equation}
where $I$ is the microinertia per unit fluid mass and $\vb*{\omega}$ is the internal rotation of the fluid elements. Combining Eqs.\eqref{AMC3}-\eqref{microinertia}, we obtain the equation for the internal angular momentum:
\begin{equation}
    \rho I\frac{D\vb*{\omega}}{Dt}=\rho\vb*{l}+\div\vb{\Lambda}+\vb{T}_x.
\end{equation}
Note that the stress tensor in the conventional hydrodynamics is restricted to be symmetric: $\vb{T}=\vb{T}^\mathrm{s}$, therefore, $\vb{T}_x$ is absent. This corresponds to the fact that the conventional hydrodynamics focuses only on linear momentum conservation while disregards angular momentum conservation. Thus, the antisymmetric components of the stress tensor play a crucial role in dissipation processes of angular momentum.

Following the standard approach of micropolar fluids~\cite{Eringen_1999}, we assume that the antisymmetric components of the stress tensor $\vb{T}^\mathrm{a}$ are given by
\begin{equation}
   (\vb{T}^\mathrm{a})_{ij}=\mu_r(\partial_iu_j-\partial_ju_i)-2\mu_r\epsilon_{ijk}\omega_k,
\end{equation}
with the rotational viscosity $\mu_r$. In the most simple case, the couple stress tensor and the body torque are absent, $\vb{\Lambda}=\vb*{l}=0$, we obtain the simplified equation for angular momentum:
\begin{equation}
    \rho I\frac{D\vb*{\omega}}{Dt}=4\mu_r(\vb*{\Omega}-\vb*{\omega}),
\end{equation}
which includes dissipation processes of angular momentum between the vorticity and the microrotation.

\section{Formulation}
In this section, we outline how to derive the valley hydrodynamic equations for noncentrosymmetric graphene with a staggered sublattice potential, which are correct up to the second order in electric fields. We start from the Boltzmann equation,
\begin{equation}
    \pdv{f_{\alpha\tau}}{t}+\dot{\vb*{r}}_{\alpha\tau}\vdot\pdv{f_{\alpha\tau}}{\vb*{r}}+\dot{\vb*{k}}_{\alpha\tau}\vdot\pdv{f_{\alpha\tau}}{\vb*{k}}=-\frac{f_{\alpha\tau}-f_{\alpha\tau}^{\mathrm{N}}}{\tau_{\mathrm{N}}}-\frac{f_{\alpha\tau}-f_0}{\tau_{\mathrm{R}}}-\frac{f_{\alpha\tau}-f_{\alpha-\tau}}{\tau_{\mathrm{vf}}},\label{boltzmann}
\end{equation}
where $\tau_{\mathrm{N}}$, $\tau_{\mathrm{R}}$ and $\tau_{\mathrm{vf}}$ are the relaxation times for normal (N), resistive (R), and valley flipping processes. $\tau_{\mathrm{N}}$ and $\tau_\mathrm{vf}$ stem from normal electron-electron scatterings and these processes conserve the linear momentum. On the other hand, $\tau_\mathrm{R}$ stems from Umklapp scatterings, impurity scatterings, and electron-phonon scatterings, therefore, this process does not conserve the linear momentum. The difference between N and valley flipping processes is that the former conserve the valley polarization, while the latter does not. $f_{\alpha\tau}(\vb*{r},\vb*{k},t)$ is the distribution function of an electron with band $\alpha$ and valley $\tau$. The semiclassical equations of motion read~\cite{DiXiao2010review}
\begin{equation}
    \dot{\vb*{r}}_{\alpha\tau}=\frac{1}{\hbar}\pdv{\epsilon_{\alpha\tau}(\vb*{k})}{\vb*{k}}-\dot{\vb*{k}}_{\alpha\tau}\times\vb*{\Omega}_{\alpha\tau},\quad\dot{\vb*{k}}_{\alpha\tau}=-\frac{e}{\hbar}\vb*{E},\label{SEOM}
\end{equation}
where the electric field $\vb*{E}$ is independent of position for simplicity. A more detailed explanation of other physical variables is given in the main text.

Following the standard approach~\cite{Landau_Kinetics,Lucas_2018}, the continuity equations for the charge carrier density and the linear momentum are obtained in the relaxation-time approximation as follows:
\begin{align}
    &\pdv{n}{t}+\div\vb*{j}=0,\label{pcl}\\
    &\nonumber\\
    &\pdv{P_{i}}{t}+\pdv{\Pi_{ij}}{x_j}=-enE_i-\frac{P_i}{\tau_{\mathrm{R}}},\label{mcl}
\end{align}
where $n$ and $\vb*{P}$ are the particle density and the momentum of electrons, $-en\vb*{E}$ is the driving force due to external electric fields and $\tau_{\mathrm{R}}$ is the relaxation-time for momentum-relaxing scatterings. In Eqs.~\eqref{pcl} and~\eqref{mcl}, $\vb*{j}_n$ and $\Pi_{ij}$ are the corresponding fluxes of each density. In the following, we will show the detailed derivation of Eqs.~\eqref{pcl} and~\eqref{mcl}.

\section{law of conservation of particle number}
First, integrating the Boltzmann equation~\eqref{boltzmann} over the momentum space, we obtain the following equation:
\begin{align}
    &\sum_{\alpha,\tau}\int[\dd{\vb*{k}}]\pdv{f_{\alpha\tau}}{t}+\sum_{\alpha,\tau}\int[\dd{\vb*{k}}]\dot{\vb*{r}}_{\alpha\tau}\vdot\pdv{f_{\alpha\tau}}{\vb*{r}}+\sum_{\alpha,\tau}\int[\dd{\vb*{k}}]\dot{\vb*{k}}_{\alpha\tau}\vdot\pdv{f_{\alpha\tau}}{\vb*{k}}\nonumber\\
    &\quad=-\sum_{\alpha,\tau}\int[\dd{\vb*{k}}]\frac{f_{\alpha\tau}-f_{\alpha\tau}^{\mathrm{N}}}{\tau_{\mathrm{N}}}-\sum_{\alpha,\tau}\int[\dd{\vb*{k}}]\frac{f_{\alpha\tau}-f_0}{\tau_{\mathrm{R}}}-\sum_{\alpha,\tau}\int[\dd{\vb*{k}}]\frac{f_{\alpha\tau}-f_{\alpha-\tau}}{\tau_{\mathrm{vf}}}.\label{pcl1}
\end{align}
In the following, we calculate each term respectively. The only subtle point arising in two-band systems is the treatment of the formally infinite number of particles in the filled band. In other words, we need to make a distinction between the ``particle density" and ``carrier density". However, assuming the contribution of the filled band to be constant, we can immediately see that it vanishes upon differentiation and does not contribute to the continuity equations. Integrating by parts the first term on the left-hand side of \eqref{pcl1}, we obtain the expression of the charge density $n$,
\begin{equation}
    \sum_{\alpha,\tau}\int[\dd{\vb*{k}}]\pdv{f_{\alpha\tau}}{t}=\pdv{t}\sum_{\tau}\int[\dd{\vb*{k}}][f_{+,\tau}-(1-f_{-,\tau})]=\pdv{t}(n_+-n_{-})\equiv\pdv{n}{t}.
\end{equation}
The definitions of the numbers of charge carriers in the two bands, $n_{\pm}$, are given by
\begin{subequations}
\begin{align}
    n_+&\equiv\sum_\tau\int[\dd{\vb*{k}}]f_{+,\tau},\label{n+}\\
    n_{-}&\equiv\sum_\tau\int[\dd{\vb*{k}}](1-f_{-,\tau}),\label{n-}
\end{align}
with the total carrier density being
\begin{equation}
n=n_+-n_-.
\end{equation}
Summing up the densities \eqref{n+} and \eqref{n-}, we may define the ``imbalance" or the total quasiparticle density
\begin{equation}
    n_I\equiv n_++n_-.
\end{equation}
\end{subequations}
Performing similar procedures for the second and third terms on the left-hand side of Eq.~\eqref{pcl1}, we obtain
\begin{align}
    \sum_{\alpha,\tau}\int[\dd{\vb*{k}}]\dot{\vb*{r}}_{\alpha\tau}\vdot\pdv{f_{\alpha\tau}}{\vb*{r}}
    &=\div\sum_{\tau}\int[\dd{\vb*{k}}][\dot{\vb*{r}}_{+,\tau}f_{+,\tau}-\dot{\vb*{r}}_{-,\tau}(1-f_{-,\tau})]
    \nonumber\\
    &=\div(\vb*{j}_+-\vb*{j}_{-})\equiv\div\vb*{j},
\end{align}
defining the electric current $\vb*{j}$ and the quasiparticle currents $\vb*{j}_{\pm}$:
\begin{subequations}
\begin{align}
    \vb*{j}&\equiv\vb*{j}_+-\vb*{j}_{-},\\
    \vb*{j}_{+}&=\sum_{\tau}\int[\dd{\vb*{k}}]\left(\frac{1}{\hbar}\pdv{\epsilon_{+,\tau}(\vb*{k})}{\vb*{k}}+\frac{e}{\hbar}\vb*{E}\times\vb*{\Omega}_{+,\tau}(\vb*{k})\right)f_{+,\tau},\\
    \vb*{j}_{-}&=\sum_{\tau}\int[\dd{\vb*{k}}]\left(\frac{1}{\hbar}\pdv{\epsilon_{-,\tau}(\vb*{k})}{\vb*{k}}+\frac{e}{\hbar}\vb*{E}\times\vb*{\Omega}_{-,\tau}(\vb*{k})\right)(1-f_{-,\tau}),\label{pcl3}
\end{align}
\end{subequations}
and
\begin{align}
    \sum_{\alpha,\tau}\int[\dd{\vb*{k}}]\dot{\vb*{k}}_{\alpha\tau}\vdot\pdv{f_{\alpha\tau}}{\vb*{k}}&=-\frac{e}{\hbar}\vb*{E}\vdot\sum_{\alpha,\tau}\int[\dd{\vb*{k}}]\pdv{f_{\alpha\tau}}{\vb*{k}}=-\frac{e}{\hbar}\vb*{E}\vdot\sum_{\tau}\int[\dd{\vb*{k}}]\left[\pdv{f_{+,\tau}}{\vb*{k}}-\pdv{\vb*{k}}(1-f_{-,\tau})\right]=0.\label{pcl4}
\end{align}

For the physical reasons that the total number of carriers does not change due to scatterings, we conclude that each term on the right-hand side of Eq.~\eqref{pcl1} should vanish:
\begin{equation}
    \sum_{\alpha,\tau}\int[\dd{\vb*{k}}]\frac{f_{\alpha\tau}-f_{\alpha\tau}^{\mathrm{N}}}{\tau_{\mathrm{N}}}=\sum_{\alpha,\tau}\int[\dd{\vb*{k}}]\frac{f_{\alpha\tau}-f_0}{\tau_{\mathrm{R}}}=\sum_{\alpha,\tau}\int[\dd{\vb*{k}}]\frac{f_{\alpha\tau}-f_{\alpha-\tau}}{\tau_{\mathrm{vf}}}=0.\label{pcl5}
\end{equation}
Summarizing the above equations~\eqref{pcl1}-\eqref{pcl5}, we obtain the continuity equation~\eqref{pcl},
\begin{equation}
    \pdv{n}{t}+\div\vb*{j}=0.
\end{equation}

Multiplying the Boltzmann equation by $\alpha$ and integrating over all states, we can find a similar continuity equation for the imbalance density,
\begin{equation}
    \pdv{n_I}{t}+\div\vb*{j}_I=0,
\end{equation}
with the imbalance current,
\begin{equation}
    \vb*{j}_I\equiv \vb*{j}_++\vb*{j}_-.
\end{equation}
Note that we have not considered the recombination processes, which do not conserve the number of particles in each band individually. Therefore, the imbalance density is also conserved.

\section{Balance Equation of Linear Momentum}
The momentum conserving law is obtained by multiplying the Boltzmann equation~\eqref{boltzmann} with $\hbar k_i$ and integrating over the momentum space:
\begin{align}
    &\sum_{\alpha,\tau}\int[\dd{\vb*{k}}]\hbar k_i\pdv{f_{\alpha\tau}}{t}+\sum_{\alpha,\tau}\int[\dd{\vb*{k}}]\hbar k_i\dot{\vb*{r}}_{\alpha\tau}\vdot\pdv{f_{\alpha\tau}}{\vb*{r}}+\sum_{\alpha,\tau}\int[\dd{\vb*{k}}]\hbar k_i\dot{\vb*{k}}_{\alpha\tau}\vdot\pdv{f_{\alpha\tau}}{\vb*{k}}\nonumber\\
    &\quad=-\sum_{\alpha,\tau}\int[\dd{\vb*{k}}]\hbar k_i\frac{f_{\alpha\tau}-f_{\alpha\tau}^{\mathrm{N}}}{\tau_{\mathrm{N}}}-\sum_{\alpha,\tau}\int[\dd{\vb*{k}}]\hbar k_i\frac{f_{\alpha\tau}-f_0}{\tau_{\mathrm{R}}}-\sum_{\alpha,\tau}\int[\dd{\vb*{k}}]\hbar k_i\frac{f_{\alpha\tau}-f_{\alpha-\tau}}{\tau_{\mathrm{vf}}}.\label{mcl0}
\end{align}
Here and hereafter, the Einstein summation convention is implied for repeated indices. Each term on the left-hand side of Eq.~\eqref{mcl0} is calculated by taking similar procedures with particle conservation law:
\begin{align}
    \sum_{\alpha,\tau}\int[\dd{\vb*{k}}]\hbar k_i\pdv{f_{\alpha\tau}}{t}&=\pdv{t}\sum_{\alpha,\tau}\int[\dd{\vb*{k}}]\hbar k_if_{\alpha\tau}=\pdv{P_i}{t},\\
    &\nonumber\\
    \sum_{\alpha,\tau}\int[\dd{\vb*{k}}]\hbar k_i\dot{r}_{n\tau,j}\pdv{f_{\alpha\tau}}{x_j}&=\sum_{\alpha,\tau}\int[\dd{\vb*{k}}]k_i\left(\pdv{\epsilon_{\alpha\tau}(\vb*{k})}{k_j}+e\epsilon_{jkl}E_k\Omega_{n\tau,l}\right)\pdv{f_{\alpha\tau}}{x_j}\nonumber\\
    &=\pdv{x_j}\sum_{\alpha,\tau}\int[\dd{\vb*{k}}]k_i\left(\pdv{\epsilon_{\alpha\tau}(\vb*{k})}{k_j}+e\epsilon_{jkl}E_k\Omega_{n\tau,l}\right)f_{\alpha\tau}=\pdv{\Pi_{ij}}{x_j},\\
    &\nonumber\\
    \sum_{\alpha,\tau}\int[\dd{\vb*{k}}]\hbar k_i\dot{k}_{n\tau,j}\pdv{f_{\alpha\tau}}{k_j}&=-eE_j\sum_{\tau}\int[\dd{\vb*{k}}]k_i\pdv{k_j}[f_{+,\tau}-(1-f_{-,\tau})]\nonumber\\
    &=eE_i\sum_{\tau}\int[\dd{\vb*{k}}][f_{+,\tau}-(1-f_{-,\tau})]=enE_i,
\end{align}
where we have defined the total momentum $P_i$ and the momentum flux $\Pi_{ij}$ as follows:
\begin{align}
    P_i&=\sum_{\alpha,\tau}\int[\dd{\vb*{k}}]\hbar k_if_{\alpha\tau},\\
    \Pi_{ij}&=\sum_{\alpha,\tau}\int[\dd{\vb*{k}}]k_i\left(\pdv{\epsilon_{\alpha\tau}(\vb*{k})}{k_j}+e\epsilon_{jkl}E_k\Omega_{n\tau,l}\right)f_{\alpha\tau}.
\end{align}
On the other hand, each term on the right-hand side of Eq.~\eqref{mcl0} is calculated as follows:
\begin{align}
    -\sum_{\alpha,\tau}\int[\dd{\vb*{k}}]\hbar k_i\frac{f_{\alpha\tau}-f_{\alpha\tau}^{\mathrm{N}}}{\tau_{\mathrm{N}}}&=-\frac{1}{\tau_{\mathrm{N}}}\left(\sum_{\alpha,\tau}\int[\dd{\vb*{k}}]\hbar k_if_{\alpha\tau}-\sum_{\alpha,\tau}\int[\dd{\vb*{k}}]\hbar k_if_{\alpha\tau}^{\mathrm{N}}\right)=0,\\
    &\nonumber\\
    -\sum_{\alpha,\tau}\int[\dd{\vb*{k}}]\hbar k_i\frac{f_{\alpha\tau}-f_0}{\tau_{\mathrm{R}}}&=-\frac{1}{\tau_{\mathrm{R}}}\left(\sum_{\alpha,\tau}\int[\dd{\vb*{k}}]\hbar k_if_{\alpha\tau}-\sum_{\alpha,\tau}\int[\dd{\vb*{k}}]\hbar k_if_0\right)=-\frac{P_i}{\tau_{\mathrm{R}}},\\
    &\nonumber\\
    -\sum_{\alpha,\tau}\int[\dd{\vb*{k}}]\hbar k_i\frac{f_{\alpha\tau}-f_{\alpha-\tau}}{\tau_{\mathrm{vf}}}&=-\frac{1}{\tau_{\mathrm{vf}}}\left(\sum_{\alpha,\tau}\int[\dd{\vb*{k}}]\hbar k_if_{\alpha\tau}-\sum_{\alpha,\tau}\int[\dd{\vb*{k}}]\hbar k_if_{\alpha-\tau}\right)\nonumber\\
    &=-\frac{1}{\tau_{\mathrm{vf}}}\left(\sum_{\alpha,\tau}\int[\dd{\vb*{k}}]\hbar k_if_{\alpha\tau}-\sum_{\alpha,\tau}\int[\dd{\vb*{k}}]\hbar k_if_{\alpha\tau}\right)=0.
\end{align}
Here, we have used the features of the local equilibrium distribution function $f_{\alpha\tau}^{\mathrm{N}}$ and the global equilibrium distribution function $f_0$:
\begin{align}
    &\sum_{\alpha,\tau}\int[\dd{\vb*{k}}]\hbar k_if_{\alpha\tau}=\sum_{\alpha,\tau}\int[\dd{\vb*{k}}]\hbar k_if_{\alpha\tau}^{\mathrm{N}},\\
    &\sum_{\alpha,\tau}\int[\dd{\vb*{k}}]\hbar k_if_0=0.
\end{align}
After all, $P_i$ and $\Pi_{ij}$ satisfy the momentum conservation law~\eqref{mcl},
\begin{equation}
    \pdv{P_i}{t}+\pdv{\Pi_{ij}}{x_j}=-enE_i-\frac{P_i}{\tau_{\mathrm{R}}}
\end{equation}
with the driving force $-enE_i$ and the momentum-relaxing force $-P_i/\tau_{\mathrm{R}}$.

\section{Balance Equation of Valley polarization}
A necessary requirement for valleytronics applications is the ability to generate and control the valley polarization. Analogous to the spin polarization, we define the valley polarization as a population imbalance between two valleys:
\begin{equation}
    P_{\mathrm{v}}(\vb*{r},t)=\sum_\alpha n_{\alpha,+K}-\sum_\alpha n_{\alpha,-K}=\sum_{\alpha,\tau}\tau\int[\dd{\vb*{k}}]f_{\alpha\tau}(\vb*{r},\vb*{k},t).
\end{equation}
In the systems we consider, an emergent conservation law appears owing to large separation of two valleys in momentum space. The valley conservation law is obtained by multiplying the Boltzmann equation~\eqref{boltzmann} with valley index $\tau$ and integrating over the momentum space:
\begin{align}
    &\sum_{\alpha,\tau}\tau\int[\dd{\vb*{k}}]\pdv{f_{\alpha\tau}}{t}+\sum_{\alpha,\tau}\tau\int[\dd{\vb*{k}}]\dot{\vb*{r}}_{\alpha\tau}\vdot\pdv{f_{\alpha\tau}}{\vb*{r}}+\sum_{\alpha,\tau}\tau\int[\dd{\vb*{k}}]\dot{\vb*{k}}_{\alpha\tau}\vdot\pdv{f_{\alpha\tau}}{\vb*{k}}\nonumber\\
    &\quad=-\sum_{\alpha,\tau}\tau\int[\dd{\vb*{k}}]\frac{f_{\alpha\tau}-f_{\alpha\tau}^{\mathrm{N}}}{\tau_{\mathrm{N}}}-\sum_{\alpha,\tau}\tau\int[\dd{\vb*{k}}]\frac{f_{\alpha\tau}-f_0}{\tau_{\mathrm{R}}}-\sum_{\alpha,\tau}\tau\int[\dd{\vb*{k}}]\frac{f_{\alpha\tau}-f_{\alpha-\tau}}{\tau_{\mathrm{vf}}}.\label{vcl0}
\end{align}
Each term on the left-hand side of Eq.~\eqref{vcl0} is calculated by taking similar procedures with particle conservation law:
\begin{align}
    \sum_{\alpha,\tau}\tau\int[\dd{\vb*{k}}]\pdv{f_{\alpha\tau}}{t}&=\pdv{t}\sum_{\alpha,\tau}\tau\int[\dd{\vb*{k}}] f_{\alpha\tau}=\pdv{P_{\mathrm{v}}}{t},\\
    &\nonumber\\
    \sum_{\alpha,\tau}\tau\int[\dd{\vb*{k}}]\dot{\vb*{r}}_{\alpha\tau}\vdot\pdv{f_{\alpha\tau}}{\vb*{r}}&=\sum_{\alpha,\tau}\tau\int[\dd{\vb*{k}}]\left(\frac{1}{\hbar}\pdv{\epsilon_{\alpha\tau}(\vb*{k})}{\vb*{k}}+\frac{e}{\hbar}\vb*{E}\times\vb*{\Omega}_{\alpha\tau}(\vb*{k})\right)\vdot\pdv{f_{\alpha\tau}}{\vb*{r}}\nonumber\\
    &=\div\sum_{\alpha,\tau}\tau\int[\dd{\vb*{k}}]\left(\frac{1}{\hbar}\pdv{\epsilon_{\alpha\tau}(\vb*{k})}{\vb*{k}}+\frac{e}{\hbar}\vb*{E}\times\vb*{\Omega}_{\alpha\tau}(\vb*{k})\right)f_{\alpha\tau}\nonumber\\
    &=\div\vb*{j}_{\mathrm{v}},\\
    &\nonumber\\
    \sum_{\alpha,\tau}\tau\int[\dd{\vb*{k}}]\dot{\vb*{k}}_{\alpha\tau}\vdot\pdv{f_{\alpha\tau}}{\vb*{k}}&=\frac{e}{\hbar}\vb*{E}\vdot\sum_{\alpha,\tau}\tau\int[\dd{\vb*{k}}]\pdv{\vb*{k}}[f_{+,\tau}-(1-f_{-,\tau})]=0.
\end{align}
Here, we have defined the valley polarization $P_{\mathrm{v}}$ and the valley current $\vb*{j}_{\mathrm{v}}$ as follows:
\begin{align}
    P_{\mathrm{v}}&=\sum_{\alpha,\tau}\tau\int[\dd{\vb*{k}}]f_{\alpha\tau},\\
    \vb*{j}_{\mathrm{v}}&=\sum_{\alpha,\tau}\tau\int[\dd{\vb*{k}}]\left(\frac{1}{\hbar}\pdv{\epsilon_{\alpha\tau}(\vb*{k})}{\vb*{k}}+\frac{e}{\hbar}\vb*{E}\times\vb*{\Omega}_{\alpha\tau}(\vb*{k})\right)f_{\alpha\tau}.
\end{align}
On the other hand, each term on the right-hand side of Eq.~\eqref{vcl0} is calculated as follows:
\begin{align}
    -\sum_{\alpha,\tau}\tau\int[\dd{\vb*{k}}]\frac{f_{\alpha\tau}-f_{\alpha\tau}^{\mathrm{N}}}{\tau_{\mathrm{N}}}&=-\frac{1}{\tau_{\mathrm{N}}}\left(\sum_{\alpha,\tau}\tau\int[\dd{\vb*{k}}]f_{\alpha\tau}-\sum_{\alpha,\tau}\tau\int[\dd{\vb*{k}}] f_{\alpha\tau}^{\mathrm{N}}\right)=0,\\
    &\nonumber\\
    -\sum_{\alpha,\tau}\tau\int[\dd{\vb*{k}}]\frac{f_{\alpha\tau}-f_0}{\tau_{\mathrm{R}}}&=-\frac{1}{\tau_{\mathrm{R}}}\left(\sum_{\alpha,\tau}\tau\int[\dd{\vb*{k}}] f_{\alpha\tau}-\sum_{\alpha,\tau}\tau\int[\dd{\vb*{k}}] f_0\right)=-\frac{P_{\mathrm{v}}}{\tau_{\mathrm{R}}},\\
    &\nonumber\\
    -\sum_{\alpha,\tau}\tau\int[\dd{\vb*{k}}]\frac{f_{\alpha\tau}-f_{\alpha-\tau}}{\tau_{\mathrm{vf}}}&=-\frac{1}{\tau_{\mathrm{vf}}}\left(\sum_{\alpha,\tau}\tau\int[\dd{\vb*{k}}] f_{\alpha\tau}-\sum_{\alpha,\tau}\tau\int[\dd{\vb*{k}}] f_{\alpha-\tau}\right)\nonumber\\
    &=-\frac{1}{\tau_{\mathrm{vf}}}\left(\sum_{\alpha,\tau}\tau\int[\dd{\vb*{k}}] f_{\alpha\tau}+\sum_{\alpha,\tau}(-\tau)\int[\dd{\vb*{k}}]f_{\alpha-\tau}\right)\nonumber\\
    &=-\frac{1}{\tau_{\mathrm{vf}}}\left(\sum_{\alpha,\tau}\tau\int[\dd{\vb*{k}}] f_{\alpha\tau}+\sum_{\alpha,\tau}\tau\int[\dd{\vb*{k}}] f_{\alpha\tau}\right)=-\frac{2P_{\mathrm{v}}}{\tau_{\mathrm{vf}}}.
\end{align}
Here, we have used the features of the local distribution function $f_{\alpha\tau}^{\mathrm{N}}$ and the global equilibrium distribution function $f_0$:
\begin{align}
    &\sum_{\alpha,\tau}\tau\int[\dd{\vb*{k}}] f_{\alpha\tau}=\sum_{\alpha,\tau}\tau\int[\dd{\vb*{k}}] f_{\alpha\tau}^{\mathrm{N}},\\
    &\sum_{\alpha,\tau}\tau\int[\dd{\vb*{k}}] f_0=0,
\end{align}
where the second equation claims the valley polarization is absent in equilibrium. After all, $P_{\mathrm{v}}$ and $\vb*{j}_{\mathrm{v}}$ satisfy the valley conservation law,
\begin{equation}
    \pdv{P_{\mathrm{v}}}{t}+\div\vb*{j}_{\mathrm{v}}=-\frac{P_{\mathrm{v}}}{\tau_{\mathrm{R}}}-\frac{2P_{\mathrm{v}}}{\tau_{\mathrm{vf}}}.
\end{equation}

\section{Summary of continuity equations}
To summarize, we list all four continuity equations for two-band electronic systems as follows,
\begin{subequations}
\begin{gather}
    \pdv{n}{t}+\div\vb*{j}=0,\\
    \pdv{n_I}{t}+\div\vb*{j}_I=0,\\
    \pdv{P_i}{t}+\pdv{\Pi_{ij}}{x_j}=-enE_i-\frac{P_i}{\tau_\mathrm{R}},\\
    \pdv{P_\mathrm{v}}{t}+\div\vb*{j}_\mathrm{v}=-\frac{P_{\mathrm{v}}}{\tau_{\mathrm{R}}}-\frac{2P_{\mathrm{v}}}{\tau_{\mathrm{vf}}}.
\end{gather}
\end{subequations}

\section{local equilibrium distribution function}
From the above discussions, independent conserved quantities $(n,n_I,\vb*{P}_\mathrm{rel},P_\mathrm{v})$ are obtained as follows.
\begin{align}
    n(\vb*{r},t)&=\sum_{\alpha,\tau}\int[\dd\vb*{k}]f_{\alpha\tau},\\
    n_I(\vb*{r},t)&=\sum_{\alpha,\tau}\alpha\int[\dd\vb*{k}]f_{\alpha\tau},\\
    \vb*{P}(\vb*{r},t)&=\sum_{\alpha,\tau}\int[\dd\vb*{k}]\hbar(\vb*{k}+\vb*{K}_\tau)f_{\alpha\tau}=\sum_{\alpha,\tau}\int[\dd\vb*{k}]\hbar\vb*{k}f_{\alpha\tau}+\hbar\vb*{K}\sum_{\alpha,\tau}\tau\int[\dd\vb*{k}]f_{\alpha\tau},\nonumber\\
    &=\vb*{P}_{\mathrm{rel}}+\hbar\vb*{K}P_{\mathrm{v}},\\
    P_{\mathrm{v}}(\vb*{r},t)&=\sum_{\alpha,\tau}\tau\int[\dd\vb*{k}]f_{\alpha\tau},
\end{align}
with the energy and entropy densities,
\begin{align}
    E(\vb*{r},t)&=\sum_{\alpha,\tau}\int[\dd\vb*{k}]\epsilon_{\alpha\tau}(\vb*{k})f_{\alpha\tau},\\
    S(\vb*{r},t)&=-k_{\mathrm{B}}\sum_{\alpha,\tau}\int[\dd\vb*{k}]\Bigl\{f_{\alpha\tau}\log f_{\alpha\tau}+(1-f_{\alpha\tau})\log(1-f_{\alpha\tau})\Bigr\}.
\end{align}
Here, we introduce the free energy which is a functional of non-equilibrium distribution function $f_{\alpha\tau}$ with corresponding parameters $(\mu,\mu_I,\vb*{u},\omega_z)$ for conserved quantities $(n,n_I,\vb*{P}_\mathrm{rel},P_\mathrm{v})$,
\begin{align}
    \Omega[f_{\alpha\tau}]&=E-TS-\mu n-\vb*{u}\vdot\vb*{P}_{\mathrm{rel}}-\hbar\omega_zP_{\mathrm{v}}\nonumber\\
    &=\sum_{\alpha\tau}\int[\dd\vb*{k}]\biggl[\Bigl\{\epsilon_{\alpha\tau}(\vb*{k})-\mu-\alpha\mu_I-\vb*{u}\vdot\hbar\vb*{k}-\hbar\omega_z\tau\Bigr\}f_{\alpha\tau}+k_{\mathrm{B}}T\Bigl\{f_{\alpha\tau}\log f_{\alpha\tau}+(1-f_{\alpha\tau})\log(1-f_{\alpha\tau})\Bigr\}\biggr].
\end{align}
The local equilibrium distribution function is obtained by the variational principle for the functional $\Omega[f_{\alpha\tau}]$,
\begin{equation}
    \left.\frac{\delta\Omega}{\delta f_{\alpha\tau}}\right|_{f_{\alpha\tau}=f_{\alpha\tau}^{\mathrm{N}}}=\epsilon_{\alpha\tau}(\vb*{k})-\mu-\alpha\mu_I-\vb*{u}\vdot\hbar\vb*{k}-\hbar\omega_z\tau+k_{\mathrm{B}}T\log\frac{f_{\alpha\tau}^{\mathrm{N}}}{1-f_{\alpha\tau}^{\mathrm{N}}}=0.
\end{equation}
In this end, we obtain the local equilibrium distribution function:
\begin{equation}
    f_{\alpha\tau}^{\mathrm{N}}=\left[1+\exp\left\{\frac{\epsilon_{\alpha\tau}-\hbar\vb*{k}\vdot\vb*{u}-\hbar\omega_z\tau-\mu_\alpha}{k_\mathrm{B}T}\right\}\right]^{-1},\label{localdist}
\end{equation}
where $\mu_\alpha\equiv\mu+\alpha\mu_I$ is the local chemical potential, $\vb*{u}$ is the hydrodynamic velocity, and $\omega_z$ is the microrotation. In the following analysis, we assume that $\mu_I=0$ for simplicity. In global equilibrium, the distribution function is given by
\begin{equation}
    f_0(\vb*{k})=f_{\alpha\tau}^\mathrm{N}(\vb*{k})|_{\mu_I=0,\vb*{u}=0,\omega_z=0}.
\end{equation}

\section{model hamiltonian}
In order to use the local distribution function Eq.\eqref{localdist} in practical calculations, we need to specify the quasiparticle spectrum. A monolayer graphene with a staggered sublatice potential breaking the inversion symmetry is a concrete example for considering valley hydrodynamics. The effective Hamiltonian describing electron states in the vicinity of the $K$ and $-K$ points is given by
\begin{equation}
    H_\tau=at(\tau k_x\sigma_x+k_y\sigma_y)+\frac{\Delta}{2}\sigma_z.\label{hamiltonian}
\end{equation}
A detailed explanation of parameters is given in the main text. The energy eigenvalues are obtained as
\begin{equation}
    \epsilon_{\alpha\tau}(\vb*{k})=\alpha\sqrt{\Delta^2/4+(atk)^2}.
\end{equation}
From these energy eigenvalues, we can calculate the density of states as follows,
\begin{equation}
    D(\epsilon)=\sum_{\alpha,\tau}\int[\dd\vb{k}]\delta[\epsilon-\epsilon_{\alpha\tau}(\vb*{k})]=\frac{\abs{\epsilon}}{\pi a^2t^2}\theta(\epsilon^2-\Delta^2/4).
\end{equation}
Using the energy eigenfunctions, we also obtain the Berry curvature
\begin{equation}
    \Omega^z_{\alpha\tau}(\vb*{k})=-2\alpha\tau\frac{\Delta}{(\Delta^2+4a^2t^2k^2)^{3/2}}.
\end{equation}
The orbital magnetic moment is one of the valley contrasting parameters in 2D materials. In gapped graphene, the orbital magnetic moment is given by
\begin{equation}
    \mathfrak{m}_{\alpha\tau}^z(\vb*{k})=-\tau\frac{e}{2\hbar}\frac{2a^2t^2\Delta}{\Delta^2+4a^2t^2k^2}=\frac{e}{\hbar}\epsilon_{\alpha\tau}(\vb*{k})\Omega^z_{\alpha\tau}(\vb*{k}),
\end{equation}
which can be intuitively interpreted as the self-rotation of the Bloch wave packet.

\section{hydrodynamic variables}
In order to construct the effective hydrodynamic theory, which is correct up to first order in the velocity and the microrotation, we introduce several concepts. In hydrodynamic regime, the system reaches to local equilibrium via normal electron-electron scatterings which conserve the linear momentum and the valley polarization. For this reason, we can assume that the distribution functions are approximately described as the local equilibrium distribution function:
\begin{equation}
    f_{\alpha\tau}\simeq f_{\alpha\tau}^\mathrm{N}=\left[1+\exp\left\{\frac{\epsilon_{\alpha\tau}-\hbar\vb*{k}\vdot\vb*{u}-\hbar\omega_z\tau-\mu}{k_\mathrm{B}T}\right\}\right]^{-1}.
\end{equation}

From now on, we assume that the band energy has an isotropic parabolic dispersion with the same effective mass $m^\ast=\hbar^2\Delta/2a^2t^2$ around some valleys: $\epsilon_{\alpha\tau}=\alpha[\Delta/2+\vb*{p}^2/2m^\ast]$, where $\vb*{p}$ is defined as a deviation from the valley. This assumption is reasonable since the relevant conduction and valence bands for noncentrosymmetric graphene with a staggered sublattice potential are well described by Eq.\eqref{hamiltonian}. We also assume that the underlying effective theory is invariant under Galilean transformation: $f_{\alpha\tau}^\mathrm{N}(\vb*{p}+\alpha m^\ast\vb*{u})=f_0(\epsilon_{\alpha\tau}(\vb*{p}))$ for an free-like dispersion $\epsilon_{\alpha\tau}(\vb*{p})=\alpha[\Delta/2+\vb*{p}^2/2m^\ast]$.

\subsection{I. Law of Conservation of particle number}
We are now ready to express three conservation laws in terms of hydrodynamic variables. First, the quasiparticle densities $n_\pm$ can be expressed as follows:
\begin{subequations}
\begin{align}
    n_+&=\sum_{\tau}\int[\dd{\vb*{k}}]f_{+,\tau}^{\mathrm{N}}(\vb*{k})=\sum_{\tau}\int[\dd{\vb*{p}}]f_{+,\tau}^{\mathrm{N}}(\vb*{p}+m^\ast_+\vb*{u})=\sum_\tau\int[\dd{\vb*{k}}]f_0(\epsilon_{+,\tau}(\vb*{k})-\hbar\tau\omega_z)\nonumber\\
    &=\sum_\tau\int[\dd{\vb*{k}}]f_0(\epsilon_{+,\tau}(\vb*{k}))+\hbar\omega_z\sum_\tau\tau\int[\dd{\vb*{k}}]\left(-\pdv{f_0(\epsilon_{+,\tau}(\vb*{k}))}{\epsilon}\right)\nonumber\\
    &=\sum_\tau\int[\dd{\vb*{k}}]f_0(\epsilon_{+,\tau}(\vb*{k})),\\
    \nonumber\\
    n_-&=\sum_{\tau}\int[\dd{\vb*{k}}][1-f_{-,\tau}^{\mathrm{N}}(\vb*{k})]=\sum_{\tau}\int[\dd{\vb*{p}}][1-f_{-,\tau}^{\mathrm{N}}(\vb*{p}+m^\ast_-\vb*{u})]=\sum_{\tau}\int[\dd{\vb*{k}}][1-f_0(\epsilon_{-,\tau}(\vb*{k})-\hbar\tau\omega_z)]\nonumber\\
    &=\sum_{\tau}\int[\dd{\vb*{k}}][1-f_0(\epsilon_{-,\tau}(\vb*{k}))]-\hbar\omega_z(\vb*{r},t)\sum_{\alpha,\tau}\tau\int[\dd{\vb*{k}}]\left(-\pdv{f_0(\epsilon_{-,\tau}(\vb*{k}))}{\epsilon}\right)\nonumber\\
    &=\sum_{\tau}\int[\dd{\vb*{k}}][1-f_0(\epsilon_{-,\tau}(\vb*{k}))],
\end{align}
where we have used the fact that the band energy $\epsilon_{\alpha\tau}(\vb*{k})$ does not depend on $\tau$. From these calculations, we obtain the charge carrier and the imbalance densities as follows:
\begin{align}
    n&=n_+-n_-=\sum_{\tau}\int[\dd{\vb*{k}}][f_0(\epsilon_{+,\tau}(\vb*{k}))-\{1-f_0(\epsilon_{-,\tau}(\vb*{k}))\}],\\
    \nonumber\\
    n_I&=n_++n_-=\sum_{\tau}\int[\dd{\vb*{k}}][f_0(\epsilon_{+,\tau}(\vb*{k}))+\{1-f_0(\epsilon_{-,\tau}(\vb*{k}))\}].
\end{align}
\end{subequations}
The same calculations can be performed for the quasiparticle currents,
\begin{subequations}
\begin{align}
    \vb*{j}_+&=\sum_{\tau}\int[\dd{\vb*{k}}]\left(\frac{1}{\hbar}\pdv{\epsilon_{+,\tau}(\vb*{k})}{\vb*{k}}+\frac{e}{\hbar}\vb*{E}\times\vb*{\Omega}_{+,\tau}(\vb*{p})\right)f_{+,\tau}^{\mathrm{N}}(\vb*{k})\nonumber\\
    &=\sum_{\tau}\int[\dd{\vb*{p}}]\left(\frac{\vb*{p}+m^\ast_+\vb*{u}}{m^\ast_+}+\frac{e}{\hbar}\vb*{E}\times\vb*{\Omega}_{+,\tau}(\vb*{p}+m^\ast_+\vb*{u})\right)\underbrace{f_{+,\tau}^{\mathrm{N}}(\vb*{p}+m^\ast_+\vb*{u})}_{f_0(\epsilon_{+,\tau}-\hbar\omega_z\tau)}\nonumber\\
    &=n_+\vb*{u}+e\hbar\omega_z\vb*{E}\times\sum_{\tau}\tau\int[\dd\vb*{p}]\Omega_{+,\tau}\left(-\pdv{f_0(\epsilon_{+,\tau})}{\epsilon}\right),\\
    \nonumber\\
    \vb*{j}_-&=\sum_{\tau}\int[\dd{\vb*{k}}]\left(\frac{1}{\hbar}\pdv{\epsilon_{-,\tau}(\vb*{k})}{\vb*{k}}+\frac{e}{\hbar}\vb*{E}\times\vb*{\Omega}_{-,\tau}(\vb*{p})\right)[1-f_{-,\tau}^{\mathrm{N}}(\vb*{k})]\nonumber\\
    &=\sum_{\tau}\int[\dd{\vb*{p}}]\left(\frac{\vb*{p}+m^\ast_-\vb*{u}}{m^\ast_-}+\frac{e}{\hbar}\vb*{E}\times\vb*{\Omega}_{-,\tau}(\vb*{p}+m^\ast_-\vb*{u})\right)[1-\underbrace{f_{-,\tau}^{\mathrm{N}}(\vb*{p}+m^\ast_-\vb*{u})}_{f_0(\epsilon_{-,\tau}-\hbar\omega_z\tau)}]\nonumber\\
    &=n_-\vb*{u}-e\hbar\omega_z\vb*{E}\times\sum_{\tau}\tau\int[\dd\vb*{p}]\Omega_{-,\tau}\left(-\pdv{f_0(\epsilon_{-,\tau})}{\epsilon}\right).
\end{align}
In these calculations, we use the fact that $\vb*{p}$ and $\vb*{\Omega}_{\alpha\tau}$ are odd under time-reversal symmetry and the relation under variable transformation $\vb*{p}\to\vb*{p}+\alpha m^\ast\vb*{u}$,
\begin{equation}
    f_{\alpha\tau}^\mathrm{N}(\vb*{p}+\alpha m^\ast\vb*{u})=f_0(\epsilon_{\alpha\tau}(\vb*{p})).
\end{equation}
From the above results, we obtain the charge and imbalance current respectively,
\begin{align}
    \vb*{j}&=\vb*{j}_+-\vb*{j}_-=n\vb*{u}+e\hbar\omega_z\vb*{E}\times\sum_{\alpha,\tau}\tau\int[\dd{\vb*{k}}]\Omega_{\alpha\tau}\left(-\pdv{f_0(\epsilon_{\alpha\tau})}{\epsilon}\right),\\
    \nonumber\\
    \vb*{j}_I&=\vb*{j}_++\vb*{j}_-=n_I\vb*{u}+e\hbar\omega_z\vb*{E}\times\sum_{\alpha,\tau}\alpha\tau\int[\dd{\vb*{k}}]\Omega_{\alpha\tau}\left(-\pdv{f_0(\epsilon_{\alpha\tau})}{\epsilon}\right).
\end{align}
\end{subequations}

\subsection{II. Balance Equation of Linear Momentum}
First we calculate the linear momentum:
\begin{align}
    P_i&=\sum_{\alpha,\tau}\int[\dd\vb*{p}]p_if_{\alpha\tau}(\vb*{p})=\sum_{\tau}\int[\dd\vb*{p}]p_i[f_{+,\tau}^{\mathrm{N}}(\vb*{p})-\{1-f_{-,\tau}^\mathrm{N}(\vb*{p})\}]\nonumber\\
    &=\sum_{\tau}\int[\dd\vb*{p}][(p_i+m_+^\ast u_i)f_0(\epsilon_{+,\tau})-(p_i+m^\ast_-u_i)\{1-f_0(\epsilon_{-,\tau})\}]\nonumber\\
    &=u_i\sum_{\tau}\int[\dd\vb*{p}][m_+^\ast f_0(\epsilon_{+,\tau})-m^\ast_-\{1-f_0(\epsilon_{-,\tau})\}]\nonumber\\
    &=m^\ast u_i\sum_{\tau}[f_0(\epsilon_{+,\tau})+\{1-f_0(\epsilon_{-,\tau})\}]\nonumber\\
    &=m^\ast u_in_I\nonumber\\
    &\equiv\rho u_i.
\end{align}
Here we have introduced the mass density $\rho=m^\ast n_I$.

Next, we would like to calculate the momentum flux,
\begin{align}
    \Pi_{ij}&=\sum_{\alpha,\tau}\int[\dd\vb*{p}]p_i\left(\pdv{\epsilon_{\alpha\tau}}{p_j}+\frac{e}{\hbar}\epsilon_{jkl}E_k\Omega_{\alpha\tau}^l\right)f_{\alpha\tau}(\vb*{p})\nonumber\\
    &=\sum_{\alpha,\tau}\int[\dd\vb*{p}](p_i+m_\alpha^\ast u_i)\left(\frac{p_j+m^\ast_\alpha u_j}{m^\ast_\alpha}+\frac{e}{\hbar}\epsilon_{jkl}E_k\Omega_{\alpha\tau}^l(\vb*{p}+m_\alpha^\ast\vb*{u})\right)f_{\alpha\tau}(\vb*{p}+m_\alpha^\ast\vb*{u})\nonumber\\
    &=p\delta_{ij}+P_iu_j.\label{S-momentum:flux}
\end{align}
Combining the particle conservation law and the linear momentum conservation law, we can derive the Euler equation correct up to $\vb*{u},\omega_z,\vb*{E}$:
\begin{equation}
    \pdv{\vb*{u}}{t}+(\vb*{u}\vdot\grad)u_i+\frac{1}{\rho}\grad p=-\frac{en\vb*{E}}{\rho}-\frac{\vb*{u}}{\tau_{\mathrm{R}}}.\label{S-Euler}
\end{equation}

\subsection{III. Balance Equation of Angular Momentum}
First, we calculate the valley polarization and the valley current respectively,
\begin{align}
    P_{\mathrm{v}}&=\sum_{\alpha\tau}\tau\int[\dd\vb*{p}]f_{\alpha\tau}(\vb*{p})=\sum_{\alpha\tau}\tau\int[\dd\vb*{p}]f_{\alpha\tau}(\vb*{p}+m^\ast_\alpha\vb*{u})=\hbar\omega_z\sum_{\alpha\tau}\int[\dd\vb*{p}]\left(-\pdv{f_0}{\epsilon}\right),\label{S-vp}\\
    \nonumber\\
    \vb*{j}_{\mathrm{v}}&=\sum_{\alpha,\tau}\tau\int[\dd\vb*{p}]\left(\pdv{\epsilon_{\alpha\tau}}{\vb*{p}}+\frac{e}{\hbar}\vb*{E}\times\vb*{\Omega}_{\alpha\tau}(\vb*{p})\right)f_{\alpha\tau}(\vb*{p})\nonumber\\
    &=\sum_{\alpha,\tau}\tau\int[\dd\vb*{p}]\left(\frac{\vb*{p}+m_\alpha^\ast\vb*{u}}{m_\alpha^\ast}+\frac{e}{\hbar}\vb*{E}\times\vb*{\Omega}_{\alpha\tau}(\vb*{p}+m^\ast_\alpha\vb*{u})\right)f_{\alpha\tau}(\vb*{p}+m^\ast_\alpha\vb*{u})\nonumber\\
    &=\vb*{u}\underbrace{\hbar\omega_z\sum_{\alpha,\tau}\int[\dd\vb*{p}]\left(-\pdv{f_0}{\epsilon}\right)}_{=P_\mathrm{v}}+\frac{e}{\hbar}\vb*{E}\times\sum_{\alpha,\tau}\tau\int[\dd\vb*{p}]\vb*{\Omega}_{\alpha\tau}f_0(\epsilon_{\alpha\tau}).\label{S-vc}
\end{align}
Here, the second term is the well-known the valley Hall effect~\cite{DiXiao2010review}. On the other hand, the first term is the longitudinal valley current $P_\mathrm{v}\vb*{u}$, which is one of our main results.

Combining the valley polarization conservation law with the explicit form of ingredients Eqs.\eqref{S-vp} and \eqref{S-vc}, we end up with the angular momentum equation:
\begin{equation}
    \pdv{\omega_z}{t}+(\vb*{u}\vdot\grad)\omega_z+\omega_z(\div\vb*{u})=-\frac{\omega_z}{\tau_{\mathrm{R}}}-\frac{2\omega_z}{\tau_{\mathrm{vf}}}.\label{S-angular}
\end{equation}

\section{Orbital Magnetization}
The thermodynamic free energy is given by
\begin{align}
    F[f_{\alpha\tau}]=-\frac{1}{\beta}\sum_{\alpha,\tau}\int[\dd\vb*{p}]\left(1+\frac{e}{\hbar}\vb*{B}\vdot\vb*{\Omega}_{\alpha\tau}(\vb*{p})\right)\log\left[1+e^{-\beta(\epsilon_{\alpha\tau}^M(\vb*{p})-\vb*{p}\vdot\vb*{u}-\tau\omega_z-\mu)}\right],
\end{align}
where $\epsilon_{\alpha\tau}^M=\epsilon_{\alpha\tau}-\vb*{B}\vdot\vb*{\mathfrak{m}}_{\alpha\tau}$ is the energy of the wave packet under magnetic fields. Then, the orbital magnetization is calculated by magnetic field derivative of the free energy,
\begin{align}
    \vb*{M}_{\mathrm{orb}}&\equiv\left.-\pdv{F}{\vb*{B}}\right|_{\vb*{B}\to0}\nonumber\\
    &=\sum_{\alpha,\tau}\int[\dd\vb*{p}]\vb*{\mathfrak{m}}_{\alpha\tau}(\vb*{p})f_{\alpha\tau}^{\mathrm{N}}(\vb*{p})+\frac{1}{\beta}\frac{e}{\hbar}\sum_{\alpha,\tau}\int[\dd\vb*{p}]\vb*{\Omega}_{\alpha\tau}(\vb*{p})\log\left[1+e^{-\beta(\epsilon_{\alpha\tau}(\vb*{p})-\vb*{p}\vdot\vb*{u}-\tau\omega_z-\mu)}\right].
\end{align}
We can see that the orbital magnetization consists of the orbital moment of carriers plus a correction from the Berry curvature. As mentioned in the main-text, the response to a perpendicular magnetic field is in fact dominated by the orbital magnetic moment. Here and hereafter, we consider 2-D inversion broken systems, therefore, the orbital magnetization has only one component along $z$-axis and its explicit form is calculated as,
\begin{align}
    M_{\mathrm{orb}}^z&\equiv\left.-\pdv{F}{\vb*{B}}\right|_{\vb*{B}\to0}\nonumber\\
    &=\sum_{\alpha,\tau}\int[\dd\vb*{p}]\mathfrak{m}^z_{\alpha\tau}(\vb*{p})f_{\alpha\tau}^{\mathrm{N}}(\vb*{p})+\frac{1}{\beta}\frac{e}{\hbar}\sum_{\alpha,\tau}\int[\dd\vb*{p}]\Omega^z_{\alpha\tau}(\vb*{p})\log\left[1+e^{-\beta(\epsilon_{\alpha\tau}(\vb*{p})-\vb*{p}\vdot\vb*{u}-\tau\omega_z-\mu}\right]\nonumber\\
    &=\hbar\omega_z\sum_{\alpha,\tau}\tau\int[\dd\vb*{p}]\left\{\mathfrak{m}^z_{\alpha\tau}\left(-\pdv{f_0(\epsilon_{\alpha\tau})}{\epsilon}\right)+\frac{e}{\hbar}\Omega^z_{\alpha\tau}f_0(\epsilon_{\alpha\tau})\right\}.
\end{align}
This result also supports that the microrotation has a meaning of an angular momentum.

\section{Introduction of viscosities}
We have derived an effective hydrodynamic theory which is composed of Eqs.\eqref{S-Euler} and \eqref{S-angular} from the Boltzmann equation and the local equilibrium distribution function, however, the obtained equations do not include the viscous effects. In order to capture these effects, we phenomelogically introduce the shear, bulk and rotational viscosities according to Ref.\cite{Lukaszewicz_1999}:
\begin{equation}
    \Pi_{ij}=\Pi_{ij}^0-\eta\left(\partial_iu_j+\partial_ju_i\right)-\lambda\delta_{ij}\partial_ku_k+\eta_r(\partial_iu_j-\partial_ju_i)-2\eta_r\epsilon_{ijk}\omega_k.
\end{equation}
Here, $\Pi_{ij}^0$ is obtained in Eq.\eqref{S-momentum:flux}. Note that the assymmetric components of $\Pi_{ij}^{(\mathrm{a})}$ contribute to the relaxation of microrotation:
\begin{equation}
    \Pi_{x,i}=\epsilon_{ijk}\Pi_{jk}=\epsilon_{ijk}\Pi_{jk}^{(\mathrm{a})}=2\nu_r\left[\curl\vb*{u}-2\vb*{\omega}\right]_i.
\end{equation}
By using these phenomelogical linear momentum flux, we obtain the hydrodynamic equations:
\begin{subequations}
\begin{gather}
    \left\{\pdv{t}+(\vb*{u}\vdot\grad)\right\}\rho+\rho(\div\vb*{u})=0,\\
    \rho\left\{\pdv{t}+(\vb*{u}\vdot\grad)\right\}\vb*{u}+\grad p-(\eta+\eta_r)\Delta\vb*{u}-(\lambda+\eta-\eta_r)\grad(\div\vb*{u})-2\eta_r\curl\vb*{\omega}=-en\vb*{E}-\frac{\rho\vb*{u}}{\tau_{\mathrm{R}}},\\
    \rho I\left\{\pdv{t}+(\vb*{u}\vdot\grad)\right\}\vb*{\omega}+\rho I\vb*{\omega}(\div\vb*{u})=2\eta_r\left[\curl\vb*{u}-2\vb*{\omega}\right]-\rho I \frac{\omega_z}{\tau_\mathrm{inter}},
\end{gather}
\end{subequations}
where $\tau_\mathrm{inter}=(1/\tau_\mathrm{R}+2/\tau_\mathrm{vf})^{-1}$ is the relaxation time for intervalley scatterings. $I$ is moment of inertia per unit fluid mass. In the following analysis, we assume that the fluid is incompressible: $\div\vb*{u}=0$, the above continuity equations become as follows:
\begin{subequations}
\begin{align}
    \frac{D\rho}{Dt}&=0,\\
    \rho\frac{D\vb*{u}}{Dt}&=(\eta+\eta_r)\Delta\vb*{u}+2\eta_r\curl\vb*{\omega}-en\vb*{E}-\frac{\rho\vb*{u}}{\tau_{\mathrm{R}}},\\
    \rho I\frac{D{\omega}_z}{Dt}&=2\eta_r[(\curl\vb*{u})_z-2{\omega}_z]-\rho I\frac{\omega_z}{\tau_\mathrm{inter}}.
\end{align}
\end{subequations}
Here, we have introduced the convective derivative: $D/Dt\equiv\partial/\partial t+(\vb*{u}\vdot\grad)$.

\section{valley hydrodynamic generation}
In this section, we apply electric fields along the $x$-direction and consider the Poiseuille flow in gapped graphene with finite width $w$ in the $y$-direction.
\subsection{I. DC valley current}
We first consider a steady state under DC electric fields. In steady states, we set $D/Dt=0$ and obtain
\begin{align}
    \vb*{\omega}=\frac{\tau_\mathrm{eff}}{\tau_{r}}\frac{\curl\vb*{u}}{2},\qquad
    \left(\nu+\nu_r\frac{\tau_{r}}{\tau_r+\tau_\mathrm{inter}}\right)\Delta\vb*{u}=\frac{en\vb*{E}}{\rho}+\frac{\vb*{u}}{\tau_{\mathrm{R}}},\nonumber
\end{align}
where $\nu_{(r)}=\eta_{(r)}/\rho$ is the kinematic (rotational) viscosity. $\tau_r=4\nu_r/I$ and $\tau_\mathrm{eff}=(1/\tau_r+1/\tau_\mathrm{inter})$ are the rotational and the effective relaxation times. We assume that the velocity field $\vb*{u}(\vb*{r})$ has only $x$ component and $u_x$ varies in the $y$-direction. This assumption does not violate the incompressible condition: $\div\vb*{u}=0$. Under this assumption, we obtain a differential equation as follows:
\begin{equation}
    \left[1-\ell^2\dv[2]{y}\right]u_x(y)=-\frac{enE}{\rho},\nonumber\label{differential1}
\end{equation}
where $\ell\equiv\sqrt{\left(\nu+\nu_r\frac{\tau_r}{\tau_r+\tau_\mathrm{inter}}\right)\tau_{\mathrm{R}}}$ is a characteristic length that determines the scale of viscous effects. Taking no-slip boundary conditions $u_x(\pm w/2)=0$, the differential equation gives the following solution
\begin{equation}
    u_x(y)=-\frac{en\tau_{\mathrm{R}}}{\rho}\left[1-\frac{\cosh(y/\ell)}{\cosh(w/2\ell)}\right]E.
\end{equation}
The microrotation $\omega_z$ corresponds to the vorticity $(\curl\vb*{u})_z/2$ and calculated as
\begin{equation}
    \omega_z=-\frac{1}{2}\frac{\tau_\mathrm{eff}}{\tau_r}\pdv{u_x(y)}{y}=-\frac{en\tau_\mathrm{R}}{2\rho\ell}\frac{\tau_\mathrm{eff}}{\tau_r}\frac{\sinh(y/\ell)}{\cosh(w/2\ell)}E.
\end{equation}
Therefore, we obtain the valley polarization and the nonlinear longitudinal valley current respectively:
\begin{subequations}
\begin{align}
    P_\mathrm{v}(y)&=\hbar\omega_z(y)D(\mu)=-\frac{\hbar D(\mu)}{2\ell}\frac{en\tau_\mathrm{R}}{\rho}\frac{\tau_\mathrm{eff}}{\tau_r}\frac{\sinh(y/\ell)}{\cosh(w/2\ell)}E,\\
    j_{\mathrm{v},\parallel}(y)&=P_\mathrm{v}(y)u_x(y)=\frac{\hbar D(\mu)}{2\ell}\left(\frac{en\tau_\mathrm{R}}{\rho}E\right)^2\frac{\tau_\mathrm{eff}}{\tau_r}\frac{\sinh(y/\ell)}{\cosh(w/2\ell)}\left[1-\frac{\cosh(y/\ell)}{\cosh(w/2\ell)}\right].
\end{align}
\end{subequations}
Notably, these results indicate that the rotational viscosity $\nu_r$ is necessary for realizing the longitudinal valley current.

\subsection{II. AC valley current}
We assume that deviations of the local thermodynamic parameters from their equilibrium values are small to justify the use of linearized hydrodynamic equations. The AC valley current is obtained by looking for a solution in the form of plane waves, i.e., $\vb*{E}=\Re[\tilde{\vb*{E}}e^{-i\Omega t}]$ together with similar expressions for other oscillating variables: $n(\vb*{r},t)=n_0+\Re[\tilde{n}(\vb*{r})e^{-i\Omega t}]$, $\rho(\vb*{r},t)=\rho_0+\Re[\tilde{\rho}(\vb*{r})e^{-i\Omega t}]$, $\vb*{u}(\vb*{r},t)=\Re[\tilde{\vb*{u}}(\vb*{r})e^{-i\Omega t}]$, and $\omega_z(\vb*{r},t)=\Re[\tilde{{\omega}}_z(\vb*{r})e^{-i\Omega t}]$. Substituting these variables into the incompressible hydrodynamic equations, we obtain
\begin{subequations}
\begin{align}
    -i\Omega\tilde{\rho}&=0,\label{ac1}\\
    -i\Omega\tilde{\vb*{u}}&=(\nu+\nu_r)\Delta\tilde{\vb*{u}}+2\nu_r\curl\tilde{\vb*{\omega}}-\frac{en_0\tilde{\vb*{E}}}{\rho_0}-\frac{\tilde{\vb*{u}}}{\tau_{\mathrm{R}}},\label{ac2}\\
    -i\Omega\tilde{{\omega}}_z&=\frac{1}{\tau_r}\left[\frac{(\curl\tilde{\vb*{u}})_z}{2}-\tilde{{\omega}}_z\right]-\frac{\tilde{\omega}_z}{\tau_\mathrm{inter}}\label{ac3}.
\end{align}
\end{subequations}
We calculate the microrotation from Eq.\eqref{ac3} as
\begin{equation}
    \tilde{{\omega}}_z=\frac{1}{1-i\Omega \tau_\mathrm{eff}}\frac{\tau_\mathrm{eff}}{\tau_r}\frac{(\curl\tilde{\vb*{u}})_z}{2}.\label{ac4}
\end{equation}
Substituting Eq.\eqref{ac4} into Eq.\eqref{ac2}, Eq.\eqref{ac2} becomes
\begin{align}
    -i\Omega\tilde{\vb*{u}}=\left(\nu+\nu_r\frac{1-i\Omega\tau_\mathrm{inter}}{1-i\Omega\tau_\mathrm{eff}}\frac{\tau_r}{\tau_r+\tau_\mathrm{inter}}\right)\Delta\tilde{\vb*{u}}-\frac{en_0\tilde{\vb*{E}}}{\rho_0}-\frac{\tilde{\vb*{u}}}{\tau_{\mathrm{R}}},\nonumber
\end{align}
where we have used the identity $\curl(\curl\tilde{\vb*{u}})=\grad(\div\tilde{\vb*{u}})-\laplacian\tilde{\vb*{u}}=-\laplacian\tilde{\vb*{u}}$ under the incompressive condition: $\div\tilde{\vb*{u}}=0$. We obtain the similar differential equation to Eq.\eqref{differential1} as,
\begin{align}
    \left[1-\tilde{\ell}^2(\omega)\dv[2]{y}\right]\tilde{u}_x(y)=-\frac{en_0\tau_{\mathrm{R}}}{\rho_0}\frac{\tilde{E}}{1-i\Omega\tau_{\mathrm{R}}},\nonumber
\end{align}
where we have introduced the effective length $\tilde{\ell}(\omega)\in\mathbb{C}$:
\begin{equation}
    \tilde{\ell}(\omega)\equiv\sqrt{\left(\nu+\nu_r\frac{1-i\Omega\tau_\mathrm{inter}}{1-i\Omega\tau_\mathrm{eff}}\frac{\tau_r}{\tau_r+\tau_\mathrm{inter}}\right)\frac{\tau_\mathrm{R}}{1-i\Omega\tau_\mathrm{R}}}.
\end{equation}
This differential equation can be solved in a similar way to the case of DC electric fields,
\begin{equation}
   \tilde{u}_x(y,\omega)=-\frac{en_0\tau_\mathrm{R}}{\rho_0}\frac{\tilde{E}}{1-i\Omega\tau_\mathrm{R}}\left[1-\frac{\cosh(y/\tilde{\ell})}{\cosh(w/2\tilde{\ell})}\right]=\frac{u_x(y)|_{\ell\to\tilde{\ell}(\omega)}}{1-i\Omega\tau_\mathrm{R}}.
\end{equation}
Then, the microrotation is also calculated as,
\begin{equation}
    \tilde{\omega}_z(y,\omega)=-\frac{1}{1-i\Omega\tau_\mathrm{eff}}\frac{\tau_\mathrm{eff}}{\tau_r}\frac{1}{2}\pdv{\tilde{u}_x(y)}{y}=-\frac{1}{1-i\Omega\tau_\mathrm{eff}}\frac{\tau_\mathrm{eff}}{\tau_r}\frac{en_0\tau_\mathrm{R}}{2\rho_0\tilde{\ell}}\frac{\tilde{E}}{1-i\Omega\tau_\mathrm{R}}\frac{\sinh(y/\tilde{\ell})}{\cosh(w/2\tilde{\ell})}=\frac{\omega_z(y)|_{\ell\to\tilde{\ell}(\omega)}}{(1-i\Omega\tau_\mathrm{eff})(1-i\Omega\tau_\mathrm{R})}.
\end{equation}
Therefore, the AC longitudinal valley current is given by,
\begin{align}
    j_{\mathrm{v},\parallel}^{\mathrm{AC}}(y,t)=P_{\mathrm{v}}(y,t)u_x(y,t)
    &\propto\Re[\tilde{\omega}_z(y,\omega)e^{-i\Omega t}]\Re[\tilde{u}_x(y,\omega)e^{-i\Omega t}]\nonumber\\
    &=\frac{1}{4}(\tilde{\omega}_z\tilde{u}_x^\ast+\mathrm{c.c.})+\frac{1}{4}(\tilde{\omega}_z\tilde{u}_xe^{-2i\Omega t}+\mathrm{c.c.})\nonumber\\
    &=\frac{1}{2}\Re[\tilde{\omega}_z\tilde{u}_x^\ast]+\frac{1}{2}\Re[\tilde{\omega}_z\tilde{u}_xe^{-2i\Omega t}].\nonumber
\end{align}
As can be seen, the nonlinear longitudinal valley current has two components correspond to the valley counterparts of the rectification and the second harmonic generation:
\begin{subequations}
\begin{gather}
j_{\mathrm{v},\parallel}^{\mathrm{AC}}(y,t)=j_{\mathrm{v},\parallel}^0(y)+j_{\mathrm{v},\parallel}^{2\Omega}(y,t),\\
j_{\mathrm{v},\parallel}^0(y)\propto\frac{1}{2}\Re[\tilde{\omega}_z\tilde{u}_x^\ast],\\
j_{\mathrm{v},\parallel}^{2\Omega}(y,t)\propto\frac{1}{2}\Re[\tilde{\omega}_z\tilde{u}_xe^{-2i\Omega t}].
\end{gather}
\end{subequations}

\section{Circular photovalley generation}
Finally, we consider bulk systems with normally-incident circularly polarized light (CPL). CPL with the electric component $\vb*{E}(t)=E_0(\cos\Omega t,\xi\sin\Omega t)$ induces a circular motion of electrons, which in tern generates a DC orbital magnetization,
\begin{equation}
{M}_{\mathrm{orb}}^z=-\frac{ne^3}{4{m^\ast}^2\Omega^3}\xi E_0^2=-\mathrm{sgn}(\mu)\frac{e^2}{4\hbar}\frac{e}{2\pi}\frac{a^2t^2}{(\hbar\Omega)^3}\left(\frac{\mu^2-\Delta^2/4}{\Delta^2/4}\right)\xi E_0^2\theta(\abs{\mu}-\Delta/2),\label{eq:m_orb}
\end{equation}
where the different chirality indices $\xi=\pm1$ correspond to the clockwise/counterclockwise circular polarizations and $\theta(x)$ denotes the Heavyside step function. This phenomenon is known as the inverse Faraday effect owing to the fact that CPL has a spin angular momentum defined as
\begin{equation}
    \vb*{S}=\frac{1}{4\Omega}\Im[\epsilon_0\tilde{\vb*{E}}^\ast\times\tilde{\vb*{E}}+\mu_0\tilde{\vb*{H}}^\ast\times\tilde{\vb*{H}}],\label{eq:spin}
\end{equation}
where $\epsilon_0$ and $\mu_0$ are the dielectric
permittivity and the permeability of vacuum. When the electromagnetic fields propagate through a vacuum, the electric and the magnetic components have the same contribution to Eq.\eqref{eq:spin}, therefore, the spin angular momentum of CPL can be effectively described only by electric fields:
\begin{equation}
    \vb*{S}=\frac{1}{2\Omega}\Im[\epsilon_0\tilde{\vb*{E}}^\ast\times\tilde{\vb*{E}}]=\frac{\epsilon_0}{2\Omega}\xi E_0^2\hat{\vb*{z}}.\nonumber
\end{equation}
As a result, the CPL-induced orbital magnetization can be rewritten in terms of the spin angular momentum as,
\begin{equation}
    {M}_{\mathrm{orb}}^z=-\frac{ne^3}{2\epsilon_0{m^\ast}^2\Omega^2}{S}^z.\nonumber\\
\end{equation}

On the other hand, our hydrodynamic theory reveals that the orbital magnetization is described as,
\begin{equation}
M_{\mathrm{orb}}^z(\vb*{r},t)=\hbar\omega_z\sum_{\alpha,\tau}\tau\int[\dd\vb*{p}]\left\{\mathfrak{m}^z_{\alpha\tau}\left(-\pdv{f_0(\epsilon_{\alpha\tau})}{\epsilon}\right)+\frac{e}{\hbar}\Omega^z_{\alpha\tau}f_0(\epsilon_{\alpha\tau})\right\}=-\frac{e}{2\pi}\omega_z\theta(\abs{\mu}-\Delta/2).\label{eq:m_orb_hydro}
\end{equation}
Comparing Eq.\eqref{eq:m_orb} with Eq.\eqref{eq:m_orb_hydro}, the inverse Faraday effect can be regarded as a direct transfer mechanism of angular momentum from CPL to the microrotation:
\begin{equation}
\omega_z^{\mathrm{inject}}=\mathrm{sgn}(\mu)\frac{e^2}{4\pi}\frac{a^2t^2}{(\hbar\Omega)^3}\left(\frac{\mu^2-\Delta^2/4}{\Delta^2/4}\right)\xi E_0^2.
\end{equation}

We are now ready to discuss the generation of a valley polarization by CPL. In the following analysis, we outline how to obtain a DC nonlinear valley polarization. We start from the hydrodynamic equations with an external angular momentum injection term stems from the inverse Faraday effect:
\begin{subequations}
\begin{gather}
\pdv{\vb*{u}}{t}+(\vb*{u}\vdot\nabla)\vb*{u}=(\nu+\nu_r)\laplacian\vb*{u}+2\nu_r\curl\vb*{\omega}-\frac{en\vb*{E}}{\rho}-\frac{\vb*{u}}{\tau_\mathrm{R}},\label{Eq:velocity}\\
\pdv{\omega_z}{t}+(\vb*{u}\vdot\nabla)\omega_z=\frac{1}{\tau_r}\left[\frac{(\curl\vb*{u})_z}{2}-\omega_z\right]-\frac{\omega_z}{\tau_\mathrm{inter}}+g\Im[\tilde{\vb*{E}}^\ast\times\tilde{\vb*{E}}]_z.\label{Eq:micro}
\end{gather}\label{Eq:CPVG}
\end{subequations}
The phenomelogically-introduced term $g\Im[\tilde{\vb*{E}}^\ast\times\tilde{\vb*{E}}]_z=g\xi E_0^2$ is the second order in electric fields. In order to solve the nonlinear equations analytically, we first expand the velocity fields and the microrotation in series up to the second order in electric fields,
\begin{equation}
\vb*{u}=\vb*{u}^{(0)}+\vb*{u}^{(1)}+\vb*{u}^{(2)},\qquad \omega_z=\omega_z^{(0)}+\omega_z^{(1)}+\omega_z^{(2)},\qquad \vb*{u}^{(n)},\omega_z^{(n)}=O(E_0^n),\ n\in\mathbb{Z}_{\geq0}\nonumber
\end{equation}
where the subscript $(n)$ denotes the order of electric fields. Furthermore, we write the solutions in Fourier components as
\begin{align}
\vb*{u}^{(0)}&=\Re[\tilde{\vb*{u}}^{(0;0)}],\qquad\vb*{u}^{(1)}=\Re[\tilde{\vb*{u}}^{(1;1)}e^{-i\Omega t}],\qquad\vb*{u}^{(2)}=\Re[\tilde{\vb*{u}}^{(2;0)}+\tilde{\vb*{u}}^{(2;2)}e^{-2i\Omega t}],\nonumber\\
\omega_z^{(0)}&=\Re[\tilde{\omega}_z^{(0;0)}],\qquad\omega_z^{(1)}=\Re[\tilde{\omega}_z^{(1;1)}e^{-i\Omega t}],\qquad\omega_z^{(2)}=\Re[\tilde{\omega}_z^{(2;0)}+\tilde{\omega}_z^{(2;2)}e^{-2i\Omega t}].\nonumber
\end{align}
Here, we should note that the second order terms include the DC and the second harmonic components. Substituting the solutions into Eqs.\eqref{Eq:CPVG}, we obtain the following equations,
\begin{align}
&\pdv{t}(\vb*{u}^{(1)}+\vb*{u}^{(2)})+\left\{\left(\vb*{u}^{(0)}+\vb*{u}^{(1)}+\vb*{u}^{(2)}\right)\vdot\nabla\right\}\left(\vb*{u}^{(0)}+\vb*{u}^{(1)}+\vb*{u}^{(2)}\right)\nonumber\\
&=(\nu+\nu_r)\laplacian\left(\vb*{u}^{(0)}+\vb*{u}^{(1)}+\vb*{u}^{(2)}\right)+2\nu_r\curl\left(\vb*{\omega}^{(0)}+\vb*{\omega}^{(1)}+\vb*{\omega}^{(2)}\right)-\frac{en\vb*{E}}{\rho}-\frac{1}{\tau_\mathrm{R}}\left(\vb*{u}^{(0)}+\vb*{u}^{(1)}+\vb*{u}^{(2)}\right),\nonumber\\
\nonumber\\
&\pdv{t}({\omega}_z^{(1)}+\omega_z^{(2)})+\left\{\left(\vb*{u}^{(0)}+\vb*{u}^{(1)}+\vb*{u}^{(2)}\right)\vdot\nabla\right\}\left(\omega_z^{(0)}+\omega_z^{(1)}+\omega_z^{(2)}\right)\nonumber\\
&=\frac{1}{\tau_r}\left[\frac{1}{2}\curl\left(\vb*{u}^{(0)}+\vb*{u}^{(1)}+\vb*{u}^{(2)}\right)-\left(\omega_z^{(0)}+\omega_z^{(1)}+\omega_z^{(2)}\right)\right]-\frac{1}{\tau_\mathrm{inter}}\left(\omega_z^{(0)}+\omega_z^{(1)}+\omega_z^{(2)}\right)+g\Im[\tilde{\vb*{E}}^\ast\times\tilde{\vb*{E}}]_z.\nonumber
\end{align}
In order to obtain the solutions, we decompose the equations according to the order of electric fields.
\begin{itemize}
\item\underline{{Zeroth order in electric fields}}

The zeroth order equations are given by,
\begin{gather}
\left(\vb*{u}^{(0)}\vdot\nabla\right)\vb*{u}^{(0)}=(\nu+\nu_r)\laplacian\vb*{u}^{(0)}+2\nu_r\curl\vb*{\omega}^{(0)}-\frac{\vb*{u}^{(0)}}{\tau_\mathrm{R}},\nonumber\\
\left(\vb*{u}^{(0)}\vdot\nabla\right)\omega_z^{(0)}=\frac{1}{\tau_r}\left[\frac{\left(\curl\vb*{u}^{(0)}\right)_z}{2}-\omega_z^{(0)}\right]-\frac{\omega_z^{(0)}}{\tau_\mathrm{inter}}.\nonumber
\end{gather}
Remembering that we have considered bulk systems, the spatial dependence of the velocity fields and the microrotation are introduced only by the non-uniformity of external electric fields. Therefore, in the zeroth order, we obtain trivial solutions:
\begin{equation}
\vb*{u}^{(0)}=0,\qquad\omega_z^{(0)}=0.
\end{equation}
These results indicate that the fluid velocity and the microrotation are absent in equilibrium.

\item\underline{{First order in electric fields}}

The first order equations are given by,
\begin{gather}
\pdv{t}\vb*{u}^{(1)}=(\nu+\nu_r)\laplacian\vb*{u}^{(1)}+2\nu_r\curl\vb*{\omega}^{(1)}-\frac{en\vb*{E}}{\rho}-\frac{\vb*{u}^{(1)}}{\tau_\mathrm{R}},\nonumber\\
\pdv{t}\omega_z^{(1)}=\frac{1}{\tau_r}\left[\frac{(\curl\vb*{u}^{(1)})_z}{2}-\omega_z^{(1)}\right]-\frac{\omega_z^{(1)}}{\tau_\mathrm{inter}}.\nonumber
\end{gather}
Here, we have used the obtained results in the above discussion: $\vb*{u}^{(0)}=\omega_z^{(0)}=0$. Normally-incident light on 2D bulk systems
does not introduce the in-plane spatial dependence of the velocity fields and the microrotation. Therefore, the above equations are rewritten as,
\begin{gather}
-i\Omega\tilde{\vb*{u}}^{(1;1)}=-\frac{en\tilde{\vb*{E}}}{\rho}-\frac{\tilde{\vb*{u}}^{(1;1)}}{\tau_\mathrm{R}},\qquad
-i\Omega \tilde{\omega}_z^{(1;1)}=-\frac{\tilde{\omega}_z^{(1;1)}}{\tau_r}-\frac{\tilde{\omega}_z^{(1;1)}}{\tau_\mathrm{inter}}.\nonumber
\end{gather}
To this end, we obtain the solutions as,
\begin{align}
\tilde{\vb*{u}}^{(1;1)}=-\frac{en\tau_\mathrm{R}}{\rho}\frac{\tilde{\vb*{E}}}{1-i\Omega\tau_\mathrm{R}},\qquad\tilde{\omega}_z^{(1;1)}=0.
\end{align}

\item\underline{{Second order in electric fields}}

The second order equations are given by,
\begin{gather}
\pdv{t}\vb*{u}^{(2)}=(\nu+\nu_r)\laplacian\vb*{u}^{(2)}+2\nu_r\curl\vb*{\omega}^{(2)}-\frac{\vb*{u}^{(2)}}{\tau_\mathrm{R}},\nonumber\\
\pdv{t}\omega_z^{(2)}=\frac{1}{\tau_r}\left[\frac{(\curl\vb*{u}^{(2)})_z}{2}-\omega_z^{(2)}\right]-\frac{\omega_z^{(2)}}{\tau_\mathrm{inter}}+g\Im[\tilde{\vb*{E}}^\ast\times\tilde{\vb*{E}}]_z,\nonumber
\end{gather}
where we have used the results that $\vb*{u}^{(1)}$ has no dependence on space and $\omega_z^{(1)}=0$. Performing similar procedures, we rewrite the above equations as,
\begin{align}
&-2i\Omega\tilde{\vb*{u}}^{(2;2)}e^{-2i\omega t}=-\frac{\tilde{\vb*{u}}^{(2;0)}}{\tau_\mathrm{R}}-\frac{\tilde{\vb*{u}}^{(2;2)}}{\tau_\mathrm{R}}e^{-2i\omega t},\nonumber\\
&-2i\Omega\tilde{\omega}_z^{(2;2)}e^{-2i\omega t}=-\frac{\tilde{\omega}_z^{(2;0)}}{\tau_r}-\frac{\tilde{\omega}_z^{(2;2)}}{\tau_r}e^{-2i\omega t}-\frac{\tilde{\omega}_z^{(2;0)}}{\tau_\mathrm{inter}}-\frac{\tilde{\omega}_z^{(2;2)}}{\tau_\mathrm{inter}}e^{-2i\omega t}+g\Im[\tilde{\vb*{E}}^\ast\times\tilde{\vb*{E}}]_z.\nonumber
\end{align}
Notably, the optical pumping term $g\Im[\tilde{\vb*{E}}^\ast\times\tilde{\vb*{E}}]$ is independent of time. Therefore, we obtain the solutions for the second order as follows,
\begin{subequations}
\begin{align}
\tilde{\vb*{u}}^{(2;0)}&=0,\qquad\tilde{\omega}_z^{(2;0)}=\tau_\mathrm{eff}g\Im[\tilde{\vb*{E}}^\ast\times\tilde{\vb*{E}}]_z,\\
\tilde{\vb*{u}}^{(2;2)}&=0,\qquad\tilde{\omega}_z^{(2;2)}=0.
\end{align}
\end{subequations}
\end{itemize}

\section{detailed calculation}
We list the detailed analytical calculations for the transport coefficients in the following,
\begin{align}
&D_\alpha(\epsilon)=\int[\dd{\vb*{k}}]\delta(\epsilon-\epsilon_{\alpha\tau}(\vb*{k}))=\frac{1}{2\pi}\frac{\abs{\epsilon}}{a^2t^2}\theta(\alpha\epsilon-\Delta/2),\\
&D(\mu)=\sum_{\alpha,\tau}D_\alpha(\mu)=\frac{1}{\pi}\frac{\abs{\mu}}{a^2t^2}\theta(\abs{\mu}-\Delta/2),\\
&n=\int\sum_{\alpha,\tau}[\dd{\vb*{k}}]f_0(\epsilon_{\alpha\tau})=\frac{\mathrm{sgn}(\mu)}{\pi}\frac{1}{2a^2t^2}(\mu^2-\Delta^2/4)\theta(\abs{\mu}-\Delta/2),\\
&\rho(\vb*{r},t)=m^\ast\sum_{\alpha,\tau}\alpha\int[\dd{\vb*{q}}]f_0(\epsilon_{\alpha\tau})=m^\ast\frac{1}{\pi}\frac{1}{2a^2t^2}(\mu^2-\Delta^2/4)\theta(\abs{\mu}-\Delta/2),\\
&\sum_{n,\tau}\tau\int[\dd{\vb*{q}}]\Omega^z_{\alpha\tau}(\vb*{q})f_0(\epsilon_{\alpha\tau}(\vb*{q}))=-\frac{\Delta}{4\pi}\left[\frac{1}{\Delta/2}\theta(\abs{\mu}-\Delta/2)-\frac{1}{\abs{\mu}}\theta(\abs{\mu}-\Delta/2)\right],\\
&\sum_{\alpha,\tau}\tau\int[\dd{\vb*{k}}]\mathfrak{m}^z_{\alpha\tau}\left(-\pdv{f_0}{\epsilon}\right)=-\frac{e\Delta}{4\pi\hbar}\frac{1}{\abs{\mu}}\theta(\abs{\mu}-\Delta/2),\\
&M^z_\mathrm{orb}=-\frac{e}{2\pi}\omega_z\theta(\abs{\mu}-\Delta/2).
\end{align}


\end{document}